\documentclass[usenatbib]{mn2e}
\usepackage{graphicx}
\usepackage{natbib}
\usepackage{aas_macros}

\def\simgt{\mathrel{\raise0.35ex\hbox{$\scriptstyle >$}\kern-0.6em
\lower0.40ex\hbox{{$\scriptstyle \sim$}}}}
\def\simlt{\mathrel{\raise0.35ex\hbox{$\scriptstyle <$}\kern-0.6em
\lower0.40ex\hbox{{$\scriptstyle \sim$}}}}

\newcommand{\Planck}{{\sl Planck}}

\newcommand{\be}{\begin{equation}}
\newcommand{\ee}{\end{equation}}
\newcommand{\bea}{\begin{eqnarray}}
\newcommand{\eea}{\end{eqnarray}}

\begin{document}

\title[SZ clusters in CMB experiments]
{Reconstructing Sunyaev-Zeldovich clusters in future CMB experiments}

\author[E.~Pierpaoli, S.~Anthoine, K.~Huffenberger, I.~Daubechies] 
{Elena Pierpaoli${}^{1}$,S.~Anthoine${}^2$, K.~Huffenberger${}^3$, I.~Daubechies${}^2$\\
${}^1$California Institute of Technology,
 Mail Code 130--33, Pasadena, CA, 91125~~USA\\
${}^2$Department of Applied Mathematics,
 Princeton University, Princeton, NJ, 08544~~USA\\
${}^3$Physics Department,
 Princeton University, Princeton, NJ, 08544~~USA\\
}

\date{Accepted ... ;
      Received ... ;
      in original form ...}

\pagerange{000--000}

\maketitle

\begin{abstract}

We present a new method for component separation
aimed to extract Sunyaev-Zeldovich (SZ) galaxy clusters from 
multifrequency maps of Cosmic  Microwave Background (CMB) experiments.
This method is designed to recover non-Gaussian, spatially 
 localized and sparse signals.
We first characterize the cluster non-Gaussianity by studying it on simulated
SZ maps.
We the apply our estimator on simulated observations  of the  
\Planck\ and Atacama Cosmology
Telescope (ACT) experiments.
The  method presented here outperforms multi-frequency Wiener filtering 
both  in the reconstructed average intensity for
given input and  in the associated error.
In the absence of point source contamination, 
 this technique reconstructs 
 the ACT (\Planck) bright (big) clusters  central $y$ parameter 
with an intensity which is  about 84 (43) 
 per cent of the original input value.
The associated error  in the reconstruction is about 12 and 27 per cent
for the 50 (12) ACT  (\Planck) clusters considered.
For ACT, the error is dominated by  beam smearing.
In the \Planck\ case the error in the reconstruction is
  largely determined by the noise level: a noise reduction by a factor 7 
would imply almost perfect reconstruction and  10 per cent error for a large 
sample of clusters.
We conclude that the selection function of \Planck\ clusters will strongly depend on the noise properties in different sky regions, as well as from the 
specific cluster extraction method assumed.
\end{abstract}


\section{Introduction}
The study of the Cosmic Microwave Background (CMB)  has greatly  improved our
understanding of the universe in the last decade.
The measurement and interpretation   of the CMB power spectrum  has determined
 the most important cosmological parameters with very high 
accuracy \citep{Spergel03}.
More experiments, now planned or underway, will produce higher resolution
multi-frequency maps of the sky in the 100--400 GHz frequency range.
One of the most important new scientific goals of these experiments is the
detection of clusters through their characteristic Sunyaev-Zeldovich (SZ)
signature \citep{SZ80}.
Because the SZ signal is substantially independent of redshift, 
SZ clusters above a mass threshold will be observed at very high distances.
Such clusters may be used to infer cosmological information
via number counts   and power spectrum
analysis  of SZ maps %. 
%Many studies have shown the great potential of this new technique 
\citep{Maju03,Levine02,Hu03,Bat03,Bat03a}.
These estimates, however, typically assume that all clusters 
above a given flux are  perfectly reconstructed and detected in the CMB maps.
In practice, this may not be the case.
SZ clusters  have radio intensities comparable to other intervening
cosmological signals like the CMB and point sources, so disentangling them is 
difficult.
%{\bf Despite the different frequency and spacial dependence of these signals, it is not so easy to disentangle them. I DON'T UNDERSTAND THIS SENTENCE, ANY WAY TO CHANGE IT ?}
Moreover, beam smearing and instrumental noise play a role in our ability
to adequately reconstruct the observed cluster. 
%We must assess how well a certain technique 
%performs in reconstructing the cluster signal given the experiment
% specifications.
For a given central comptonization parameter $y$ (see below for definition),
the reconstructed value may also depend on the cluster location and shape.
Given experimental specifications and a reconstruction technique, the 
associated reconstruction error 
needs to be assessed and  accounted for when relating cluster observables to
 cosmological models.
This error is in addition to the one usually associated with cluster
scaling relations,  which play a major role in the current determination
of $\sigma_8$ from galaxy clusters \citep{PBSW}.

%Moreover, the specific observable to use may depend on the type of experiment
%in hand. 
%In this paper we address these issues for two very different types of 
%experiment: \Planck\ and ACT.
Given the observed maps at different frequencies, clusters can be detected
with two quite different approaches:

a) a formal component separation is applied to the map, and the cluster
map is reconstructed along with the maps for all other processes. 

This approach was initially developed to reconstruct CMB maps. 
Most often it has been  applied  in Fourier space, with and without 
assumptions on the Gaussianity of the signal \citep{max96, Stoly02}. 

b) The maps at different frequencies are first combined in an optimal 
way in order to enhance the cluster signal and minimize the other ones.
A spatial filter, (typically circularly symmetric) is then applied 
to the final map \citep{Herranz02,Diego02}.

Clusters of galaxies maps present very specific features,
 in particular: 
{\it i}) clusters are ``rare'' objects in the map---they do not fill the 
majority of the space; 
{\it ii}) The cluster signal is 
non-Gaussian on several scales \citep{Zhang02,Diego03}, 
and we are most interested in the non-Gaussianity  on the scale   
of the typical cluster core size; 
{\it iii}) 
the signal on different scales is correlated.
Keeping these characteristics in mind, in 
this paper we develop a method for formal component separation
of different signals
that is tailored to
better reconstruct the SZ galaxy 
cluster signal from multifrequency maps. 
Our map reconstruction 
method is wavelet-based and it can take into account
the specific non-Gaussianity expected for a given signal (SZ clusters in
particular). We will see that the combination of these two features
 enables us to better
reconstruct the intensity of the cluster center, 
which is essential to reliably  relate the SZ signal to theoretical
quantities in order to derive cosmological parameters.

Wavelets have been applied to the analysis of multifrequency maps 
to characterize specific signals \citep{Stark04} and 
extract point sources \citep{Cayon00,Vielva01a}, to recover
the CMB sky with Maximum Entropy methods \citep{Vielva01,Maisinger04}
 and to assess statistics of the CMB \citep{Tenorio99,Hobson99,Cayon01,
Aghanim03}.
Here we focus on galaxy clusters, adopt a suitable wavelet basis, and develop
a new Bayesian estimator appropriate to the reconstruction of the non-Gaussian
 signal associated with cluster maps.

Future SZ surveys may be very different in nature. 
Space-based \Planck\ will cover the whole sky, seeing all the 
most massive clusters but
with relatively low resolution.  A number of ground-based 
experiments will cover a smaller area with 
higher resolution.    Because
the performance of the reconstruction method also depends on the experimental 
characteristics,
we apply our method to two  different  SZ experiments,
 the \Planck\ surveyor and the Atacama Cosmology Telescope (ACT).
The main purpose of this paper is to introduce our new estimator,
on which component separation is based, 
 and to test  and compare it with standard techniques.
After reconstructing the SZ maps, we  assess which error is involved 
in the determination  of SZ observables from the map.
The analysis of simulated maps allows us to design observables  
suited to derive cosmological parameters for a given experiment.
We will also assess the impact  of instrumental limitations 
 in recovering the cluster maps.
We then  assess the level of completeness and purity of the surveys
for different flux cuts.
Throughout the paper we compare our method with the standard multi-frequency 
Wiener filtering technique applied in the same wavelet 
space.

 This paper is organized as follows:
in section \ref{par:signals} we describe the relevant signals 
at radio/infrared frequencies and the characteristics of the experiments 
that we are going to consider; we then describe our wavelet basis and 
reconstruction method in section \ref{par:recmet}; section \ref{par:res}
 is dedicated to the results and section \ref{par:con} to the conclusions.

\section{Astrophysical signals and instrument characteristics} 
\label{par:signals}

We will consider   experiments that will provide a map of the sky 
in the frequency range 100--400 GHz.
In this range we expect to observe several galactic and extragalactic signals,
 like the synchrotron  and dust emission from the Galaxy,
 the CMB, radio and infrared point sources and SZ galaxy clusters.
We are interested here in the reconstruction of the cluster signal.
Because the galactic signal has a very different spatial structure from
SZ clusters, we assume here that the Galaxy is not a fundamental limitation
in reconstructing SZ cluster maps, which is likely true in substantial portions
of the sky.
Point sources, particularly dusty star-forming galaxies at high-redshift
 which
shine brightly at sub-mm frequency, may be a potential concern.  The modeling of source number counts, frequency dependences, and spatial correlations remain uncertain.  We prefer to leave them out of the analysis at the
 present time, and test our technique in absence of point sources. 
For \Planck\ this approach is justified by the presence of higher and lower frequency channels, which will hopefully 
allow modeling and subtraction of point sources before 
component separation.  For ACT, which lacks these  channels,
 this is probably a too optimistic assumption
 \citep{2004ApJ...602..565W,2004astro.ph..8066H}.
However, we use it for simplicity in testing our new estimator, 
which in any case we find to be superior to Wiener filtering techniques used previously.  
We will assess in future work the contamination of such sources in 
the context of the technique presented here.

We simulate the CMB with Gaussian random fields using a power spectrum derived from the best-fitting WMAP parameters \citep{2003ApJS..148....1B}.

In the following we review the SZ cluster signal and describe the 
characteristics of the simulated maps and of 
the experiments we are discussing here. 

\subsection{Clusters of Galaxies}
CMB photons traveling from the last scattering surface to the earth 
interact with the high energy electrons in intervening massive galaxy 
clusters. As a consequence of this scattering, the CMB temperature and 
intensity is modified 
in the direction of a cluster.  This is known as the thermal and kinetic 
SZ effects.
The effect related to the electron's thermal motion causes
a change in the CMB intensity  $\delta I$ 
in the direction $\hat n$  of the cluster:
\be
{\delta I}(\hat n) = -2y(\hat n) I_0 S_I[x(\nu)],
\label{eq:deli}
\ee
  where 
\be
S_I={ {x^4 \exp(x)} \over (\exp(x) -1)^2} 
\left[ 2 - {x \over 2} { {\exp(x)+1} \over {\exp(x) - 1} } \right]
\ee
with  $x \equiv h\nu/k_B T_{CMB}= \nu / 57~{\rm GHz}$, $I_0 \equiv 2(k_B T_{CMB})^3/(hc)^2$
 and
\be
y(\hat n) = {\sigma_T \over{m_e c^2}} \int_0^{L} {n_e(l \hat n) k_B T_e(l \hat n) dl} = {\sigma_T \over{m_e c^2}}\int_0^{L} P_e(l \hat n) dl
\label{eq:yparam}
\ee
is the comptonization $y$-parameter which depends on temperature ($T_e$) 
number density ($n_e$) through the pressure ($P_e$) of free electrons.  
The integral is over the proper  distance $l$.
The corresponding temperature change reads:
\be
{\delta T_{CMB} \over T_{CMB}} = -2 y(\hat n)
\left[ 2 - {x \over 2} { {\exp(x)+1} \over {\exp(x) - 1} } \right].
\ee
Note that in the Rayleigh-Jeans limit ($x \ll 1$) we have 
$\delta T_{CMB} / T_{CMB} = \delta I_\nu / I_\nu =  -2 y(\hat n)$.
The thermal SZ signal causes a decrement (increment)  of the intensity below
 (above) the characteristic frequency of  217 GHz. 
The  thermal SZ is observable between 10 and 800 GHz,
with a minimum (maximum)  around 145 (350) GHz.
Because of this peculiar frequency dependence, multi-frequency observations
will be crucial in recovering the cluster SZ signal.
Massive clusters have typical $y$ parameters of the order of $10^{-4}$
creating an SZ signal of the same order of magnitude as that of the
 CMB fluctuations.

The interaction of free electrons in galaxy clusters with the CMB photons 
is also responsible for the  kinetic SZ.
Clusters of galaxies have a peculiar motion with respect to the Hubble flow.
As a consequence, the scattered CMB photons are subject to a Doppler effect 
due to the bulk cluster motion.
The kinetic SZ is weaker than the thermal SZ and it has a similar frequency 
dependence to the CMB, therefore it is harder to detect and separate from
 the CMB.
Very likely the kinetic SZ will be measured after cluster positions have been
 determined. We therefore ignore it in the component separation 
technique developed in the following sections.

\subsection{Characteristics of future CMB experiments}

The performance of any component separation technique also depends on the 
characteristics of the experiment in hand. Future SZ surveys will have 
a broad spectrum of possible characteristics in term of frequency coverage,
sensitivity and resolution. For this reason we specify here our analysis 
 for two future CMB experiments very different in nature:
\Planck\ and ACT\footnote{http://planck.esa.int/,  
http://www.hep.upenn.edu/act/index.html}.\nocite{2003NewAR..47..939K}
The specifications for these experiments most relevant for 
SZ surveys are summarized in 
table \ref{tab:ic}.

\Planck\ is an all-sky experiment with a quite broad beam (5 arcmin)
 if compared to 
the typical cluster size (1--10 arcmin),
therefore we expect \Planck\ to detect the most massive (or extended) 
clusters only.
The large area covered, however, will allow to detect a sizable number of
them.

ACT is a higher resolution, ground based experiment where most
 massive clusters are larger than the beam. 
At this resolution, clusters appear aspherical (see 
fig.~\ref{fig:mapsACT}, first panel)
and the challenge here is also to be able to individually detect merging 
structures and resolve the outskirts of massive clusters.
Moreover, we want to find the many small clusters which may be confused with 
noise (or other point sources).

\begin{table}
\begin{center}
\begin{tabular}{cccc}
Experiment & $\nu$ (GHz) & FWHM (arcmin) & $\sigma (\mu {\rm K})$  \\ \hline 
\hline
\Planck & 143  & 7.1  & 6  \\
-- &217  & 5.0 & 13  \\
-- & 353  & 5.0 & 40  \\
\hline
\hline
ACT &145 &1.7 & 2  \\
-- &217 &1.1 &3.3\\
-- &265 &0.93 & 4.7\\
\end{tabular}
\end{center}
\caption{ The characteristics of the \Planck\ and ACT experiments at the frequencies used in this work.  The RMS detector noise per full-width-half-maximum pixel, labeled $\sigma$, is given in thermodynamic temperature units.}
\label{tab:ic}
\end{table}

In our simulations of both experiments, the detection of the cluster central emission is
 going to be challenged by the following factors: 
{\it i)} smearing  by the instrumental  beam; {\it ii)} instrumental 
noise  {\it iii)} confusion with the CMB anisotropy structure.

In the following we assess how precise the reconstruction of the central 
emission in these experiments is.
In order to do so, we constructed simulated maps of the sky at different 
frequencies for both experiments with CMB and SZ cluster maps.
For \Planck, the  $10 \times 10$ square-degree SZ
cluster maps where taken from \cite{2003ApJ...597..650W}\footnote{http://pac1.berkeley.edu/tSZ/PlanckSZ/}.
For ACT we use $1.19 \times 1.19$ degrees maps obtained from  hydrodynamical
 simulations by \cite{Zhang02}.
For both experiments the reconstruction was based on three 
frequency channels, which are specified in table \ref{tab:ic}.

\section{The reconstruction method} \label{par:recmet}
In this section, we present the method used to reconstruct the different 
processes from the observed maps.
Instead of the usual Fourier space decomposition of the signal, we 
adopt here a wavelet decomposition.
We perform a Bayesian least square estimation of the different processes,
modeling the statistics of wavelet coefficients of the signals by Gaussian 
scale mixtures. The estimation can be thought of a  weighted local Wiener 
filters on neighborhoods of wavelet coefficients, where the weight is 
determined by the specific non-Gaussianity under consideration.
Below, we describe the details of the wavelet choice, 
and of the reconstruction method.

\subsection{The wavelet decomposition used} \label{par:wavdec}

Since we deal with sky maps, we are interested in two-dimensional wavelets.
In the choice of the wavelet to use for our image processing, 
we have to balance the orthogonality of the wavelet 
(which is desirable in order to well-define the statistics) 
with the compatibility of the wavelet with the image at hand.
Orthogonal wavelet bases, such as the 2-D Daubechies wavelets, are typically 
heavily biased towards horizontal and vertical directions; moreover, they are 
usually not very well concentrated in frequency. To avoid this, we use an overcomplete
 wavelet representation inspired by the work of  \citet{Portilla03}
 which is more adequate to the analysis of astrophysical images.

The wavelet decomposition of a signal $s$ in our case reads:
\be
s  = \sum_{q\in {Z}^2}\langle s, \phi_{q}\rangle  \phi_{q} + \sum_{j=0}^{J} 
\sum_{m=1}^{M} \sum_{q \in 2^{-j} {Z}^2} \langle s, \psi_{j,m,q}\rangle \psi_{j,m,q} 
\label{eq:wt}
\ee
where $\phi_{q}$ are the scaling functions, $\psi_{j,m,q}$ are the wavelets and
$\langle ,\rangle$ are scalar products. The sum $\sum_{q}\langle s, \phi_{q}\rangle \phi_{q}$
 is the projection of $s$ on the coarsest scale, i.e. a low-pass version of $s$.
 Each scaling coefficient $\langle s, \phi_{q}\rangle$ contains information about the signal 
$s$ at the coarsest scale and at a specific location in space $q$.
 For $j$ fixed, the sum $\sum_{m} \sum_{q} \langle s, \psi_{j,m,q}\rangle \psi_{j,m,q}$
 is the projection of $s$ on the scale $j$, i.e. a band-pass version of $s$.
 Each wavelet coefficient $\langle s, \psi_{j,m,q}\rangle$ contains information about the 
signal $s$ at the specific scale $j$, orientation $m$ and location in space $q$.
 
\begin{figure}
\begin{center}
\resizebox{\columnwidth}{!}{
\includegraphics{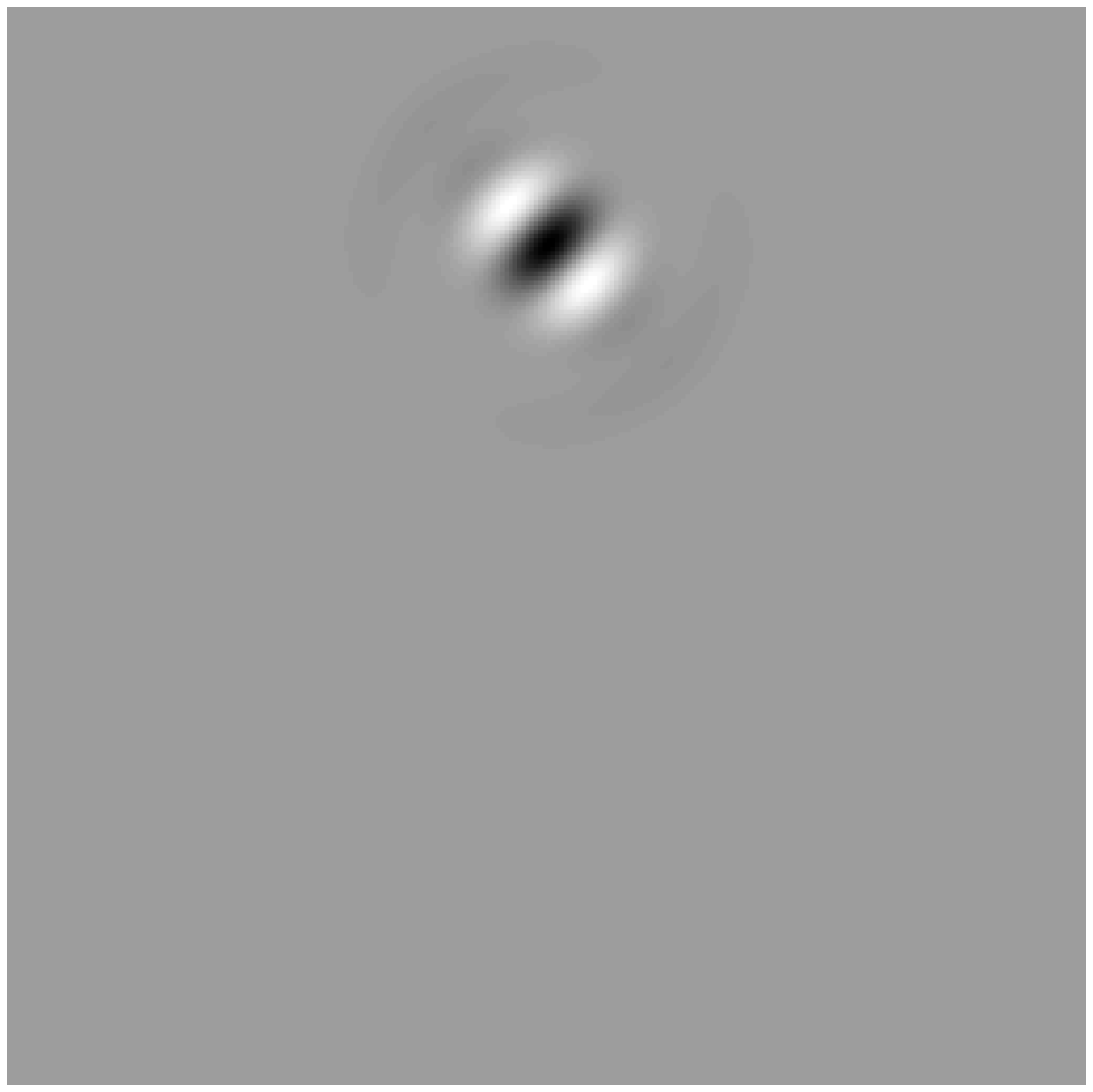}
\includegraphics{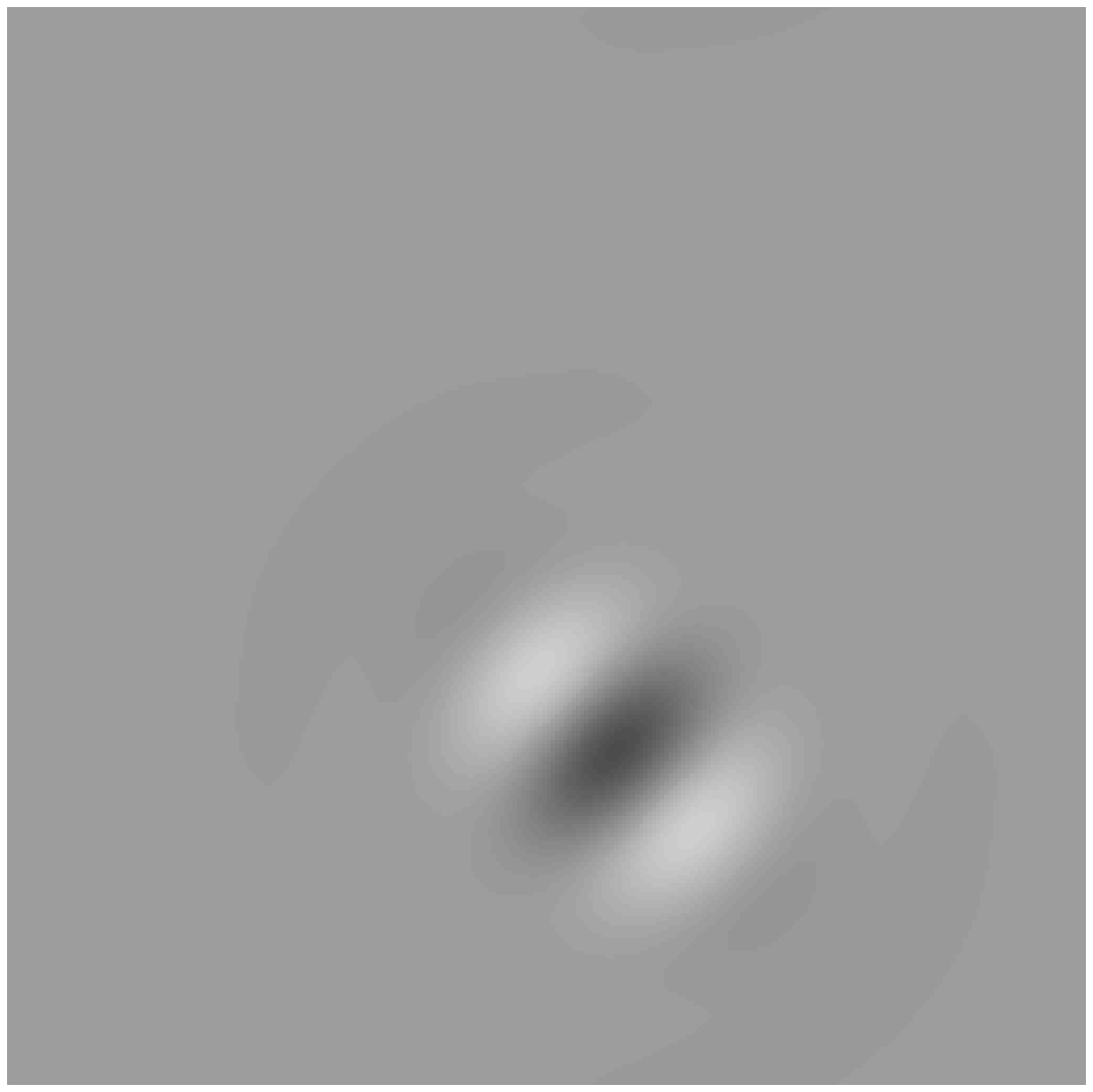}
\includegraphics{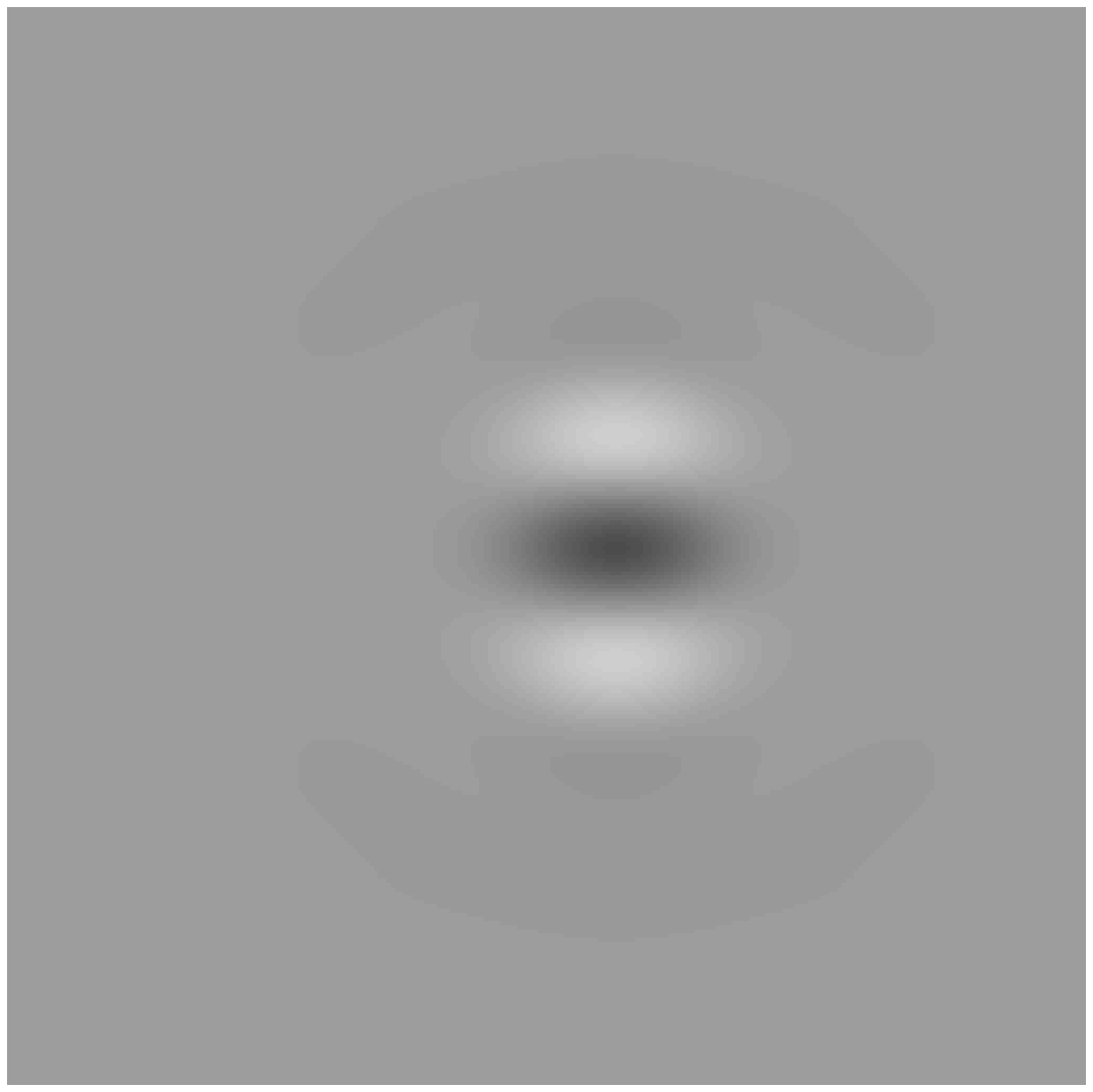}
\includegraphics{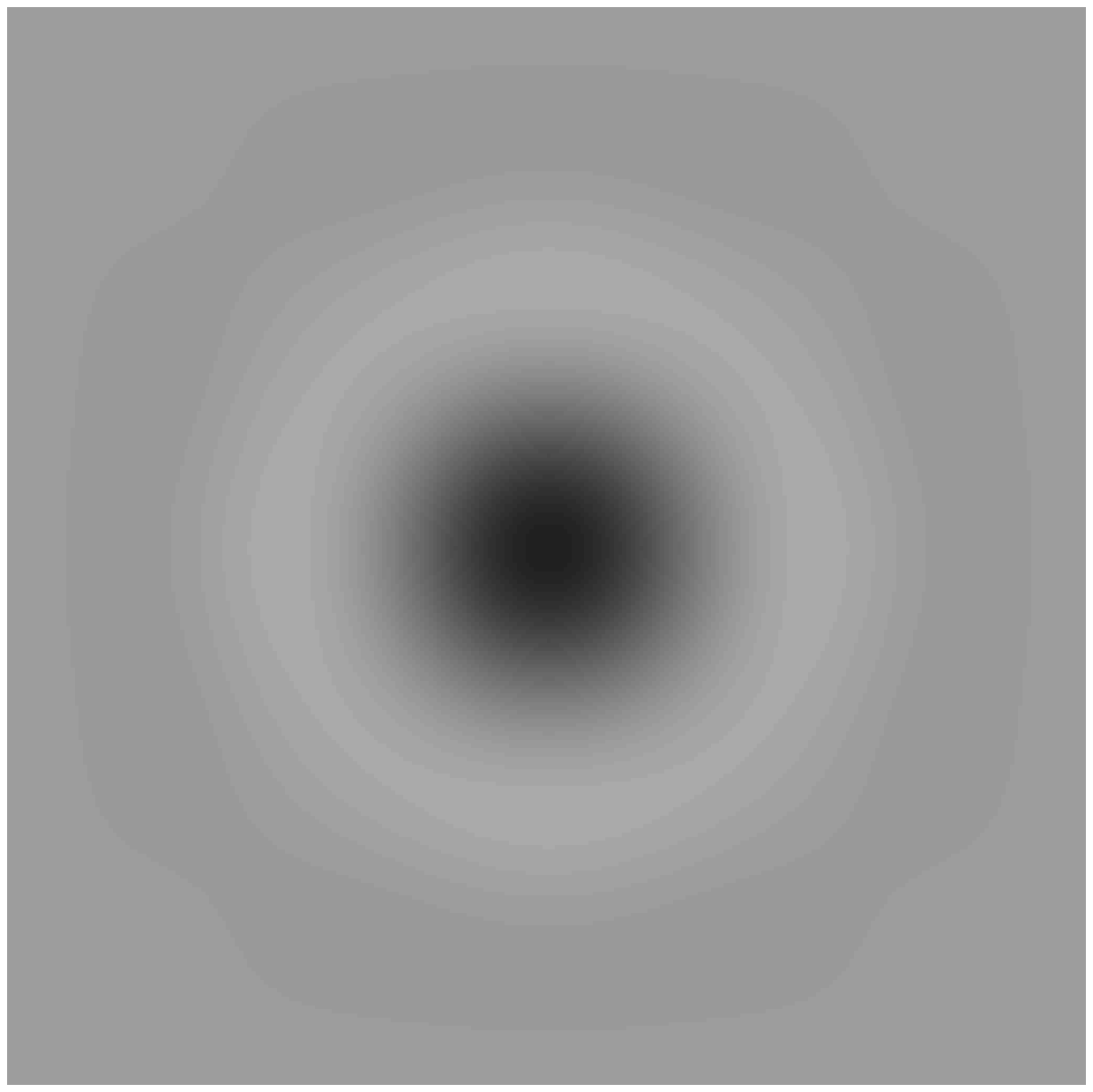}}\\
\resizebox{\columnwidth}{!}{
\includegraphics{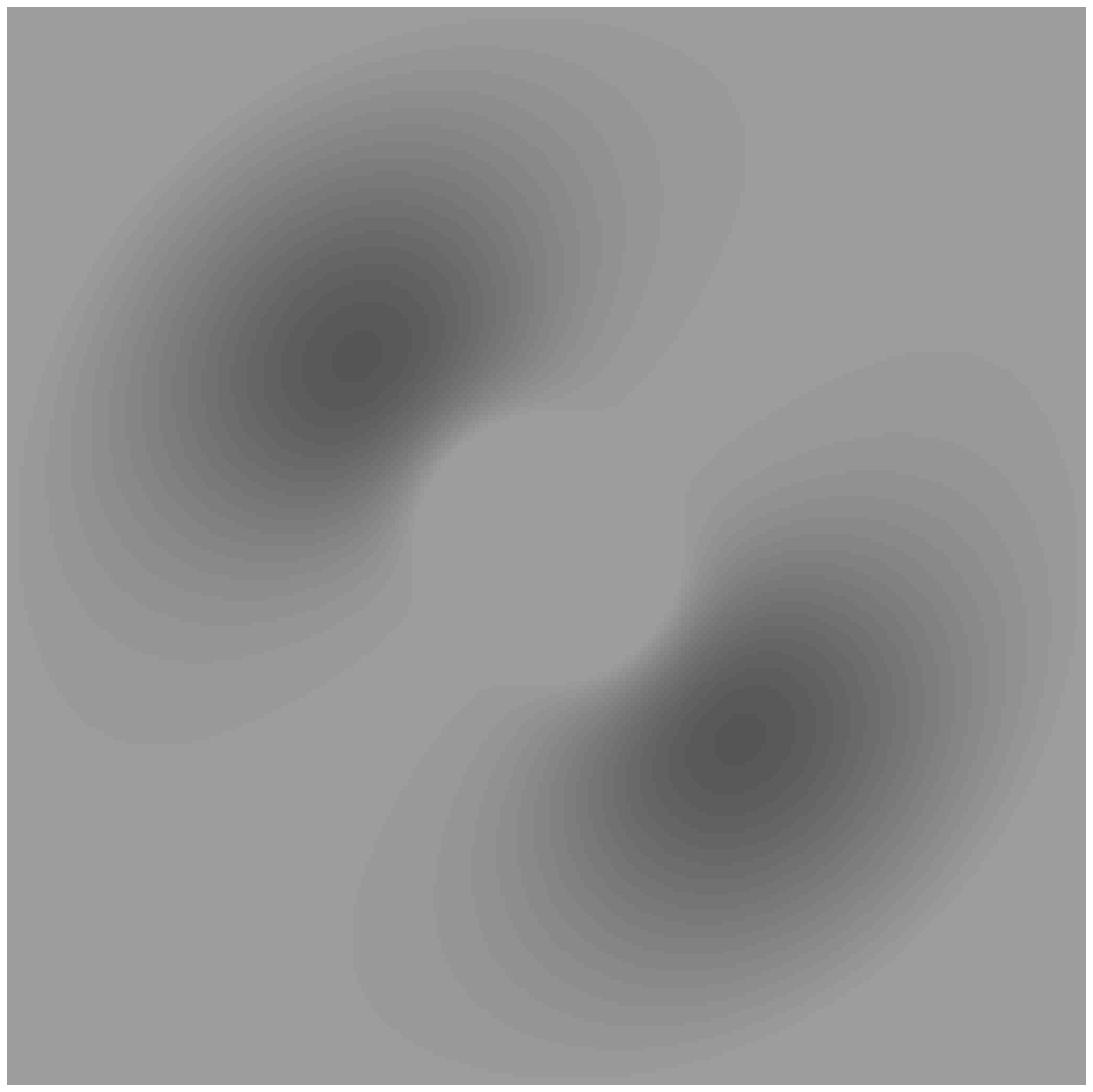}
\includegraphics{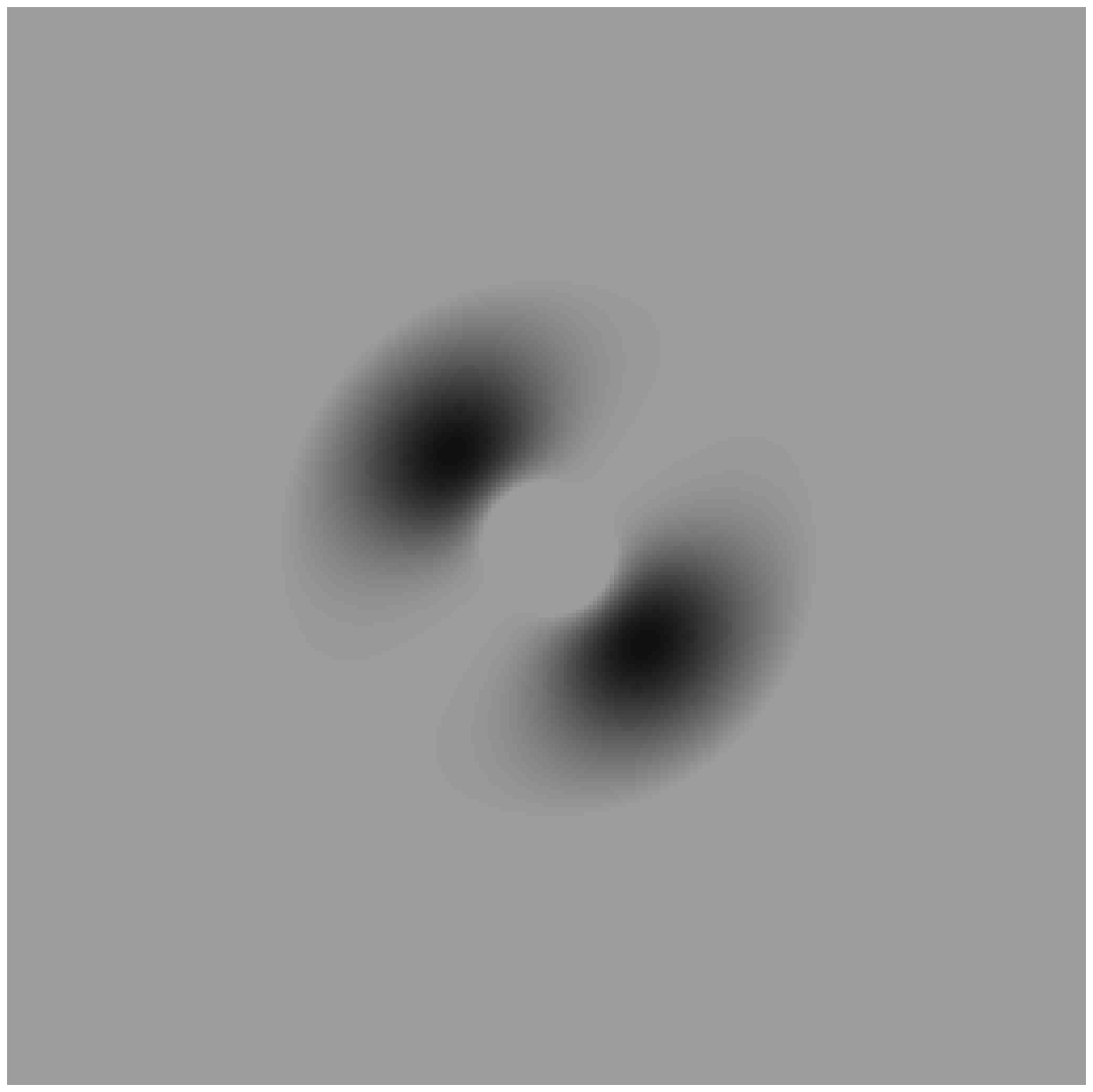}
\includegraphics{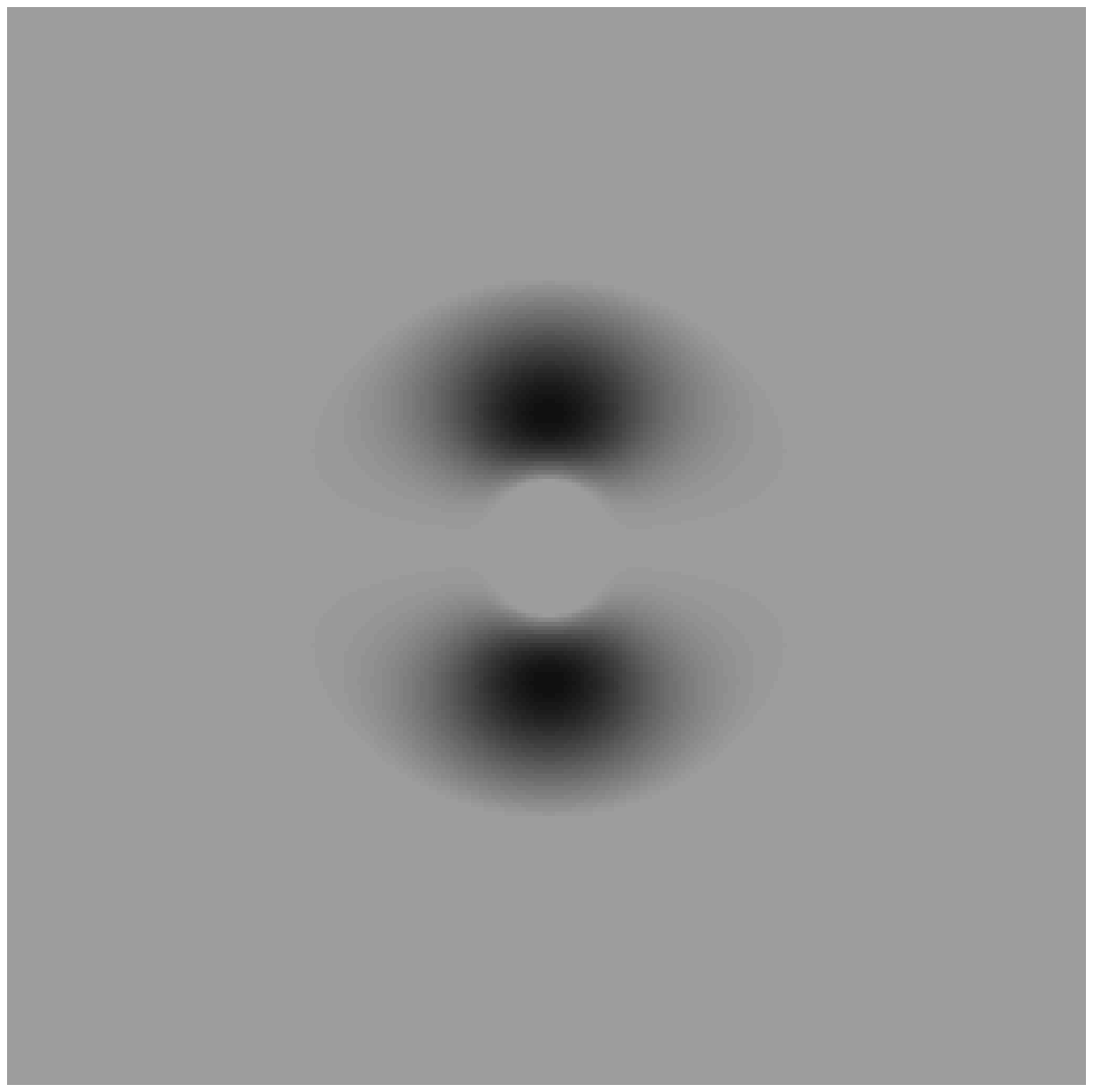}
\includegraphics{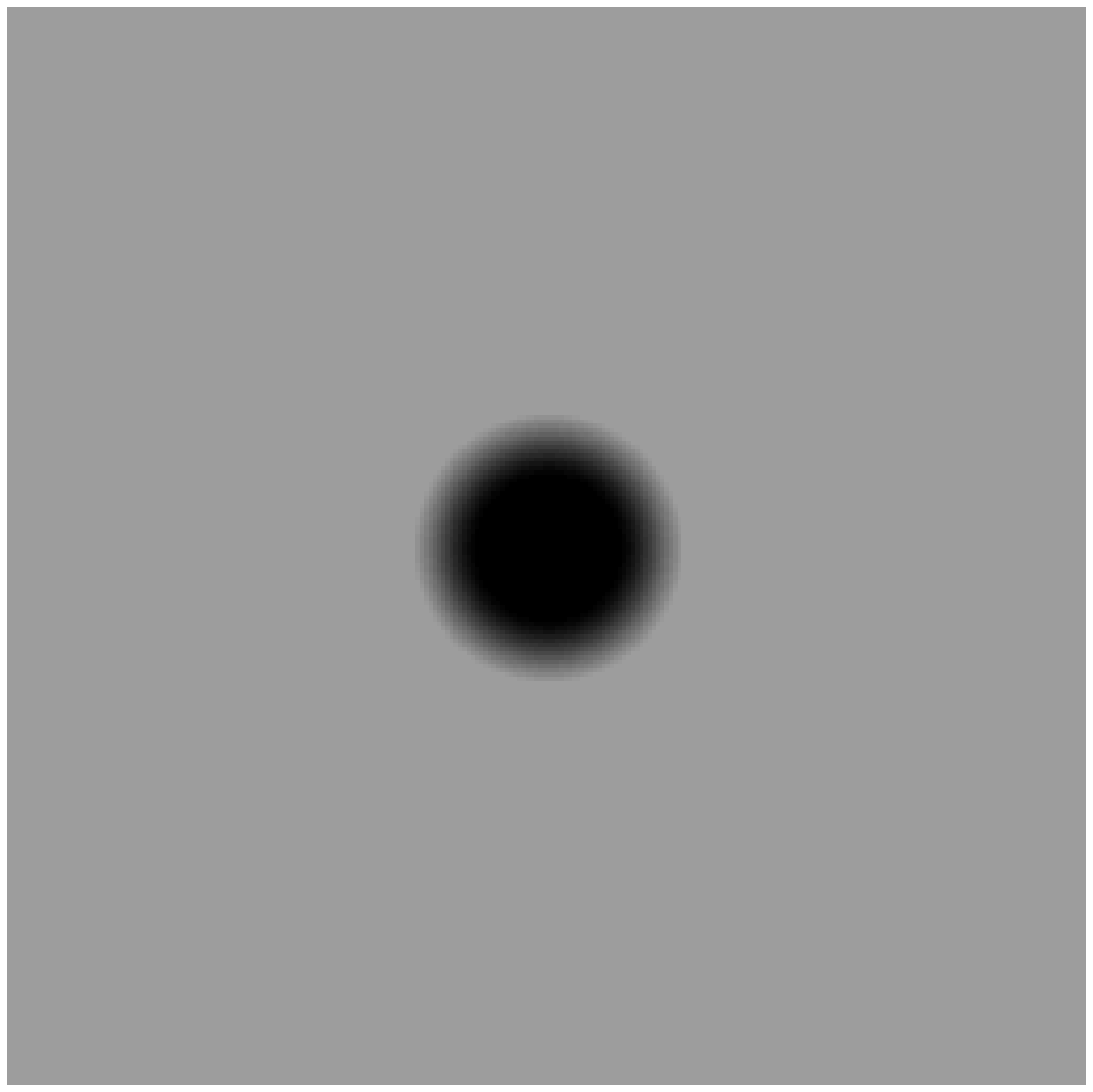}}
\end{center}
\caption{Top row: wavelets in space; Bottom row: wavelets in Fourier plane. First column: wavelet at a fine scale $j+1$, centered at location $q_0$, oriented along the first diagonal.  Second column: wavelet at a coarser scale $j$, centered at location $q_1$, oriented along the first diagonal. Third column: wavelet at the same coarser scale $j$, centered at location $q_2$, oriented along the horizontal axis.  Fourth column: scaling function, centered at location $q_2$.}
\label{fig:wav}
\end{figure}

As usual with wavelet transforms, changing scale is done by dilating, and changing 
location is done by translating the wavelet:
\be
\psi_{j+1,m,q}(x)= 2 \psi_{j,m,q}(2x) \ \ \ \ \psi_{j,m,q}(x)= \psi_{j,m,0}(x-q)
\ee
Hence, scale $j+1$ corresponds to a spatial frequency band that is twice as wide and for which the 
central frequency is twice as large as that of scale $j$. On the other hand, in space, 
the wavelets at scale $j+1$ are better localized than at scale $j$ since they are more narrowly
 concentrated around their center $q$ (see Fig.~\ref{fig:wav}, column 1 and 2). 

Unlike the 2-D Daubechies wavelets, the wavelets (and scaling function) we use here are defined
 in the Fourier plane. This ensures that they are well concentrated in frequency. Moreover, it 
enables us to introduce orientation by rotating the Fourier transform of the wavelet (see Fig.~\ref{fig:wav}, column 2 and 3). If $\tilde {f}$ is the Fourier transform of $f$ and $(r,\theta)$ are polar coordinates, then:
\be
\widetilde{\psi_{j,m,q}}(r,\theta)= \widetilde{\psi_{j,0,q}} (r,\theta-\frac{m\pi}{M})
\ee
The transform is therefore close to rotation invariant and computation is fast via FFT.

The Fourier transform of the wavelets and scaling functions are
\bea
\widetilde{\phi_{0}}(r,\theta)&=& L(2r,\theta)\\
\widetilde{\psi_{j,m,0}}(r,\theta)&=& L({r \over {2^j}}) H({{2 r} \over {2^j}})G_M (\theta - \frac{m\pi}{M})  \nonumber
\eea
for ${j\geq0,}$ and ${0\leq m<M}$, where the low-pass filter $L(r)$, the high-pass filter $H(r)$ and the
oriented filters $G_M(\theta)$ are
\bea
L(r) &=& \cos{({\pi \over 2} \log_2(r))}\delta_{1<r<2}+\delta_{r<1}  \\ \nonumber
H(r) &=& \sin{({\pi \over 2} \log_2(r))}\delta_{1<r<2}+\delta_{r>2}  \\ \nonumber
G_{M}(\theta) &=& {(M-1)! \over \sqrt{M[2(M-1)]!}}\left|2 \cos \theta\right|^{M-1} 
\eea 
The set of all wavelets and scaling functions determines a redundant system (they are linearly dependent),
 however, the Plancherel equation holds:
\be
||s||_{L^2}^2 = \sum_{q \in {Z}^2} |\langle s, \phi_{q}\rangle|^2 + \sum_{j=0}^{J} 
\sum_{m=1}^{M} \sum_{q \in 2^{-j} {Z}^2} |\langle s, \psi_{j,m,q}\rangle|^2
\ee
This ensures perfect reconstruction (eq. \ref{eq:wt}).

\begin{figure}
\begin{center}
\resizebox{\columnwidth}{!}{\includegraphics{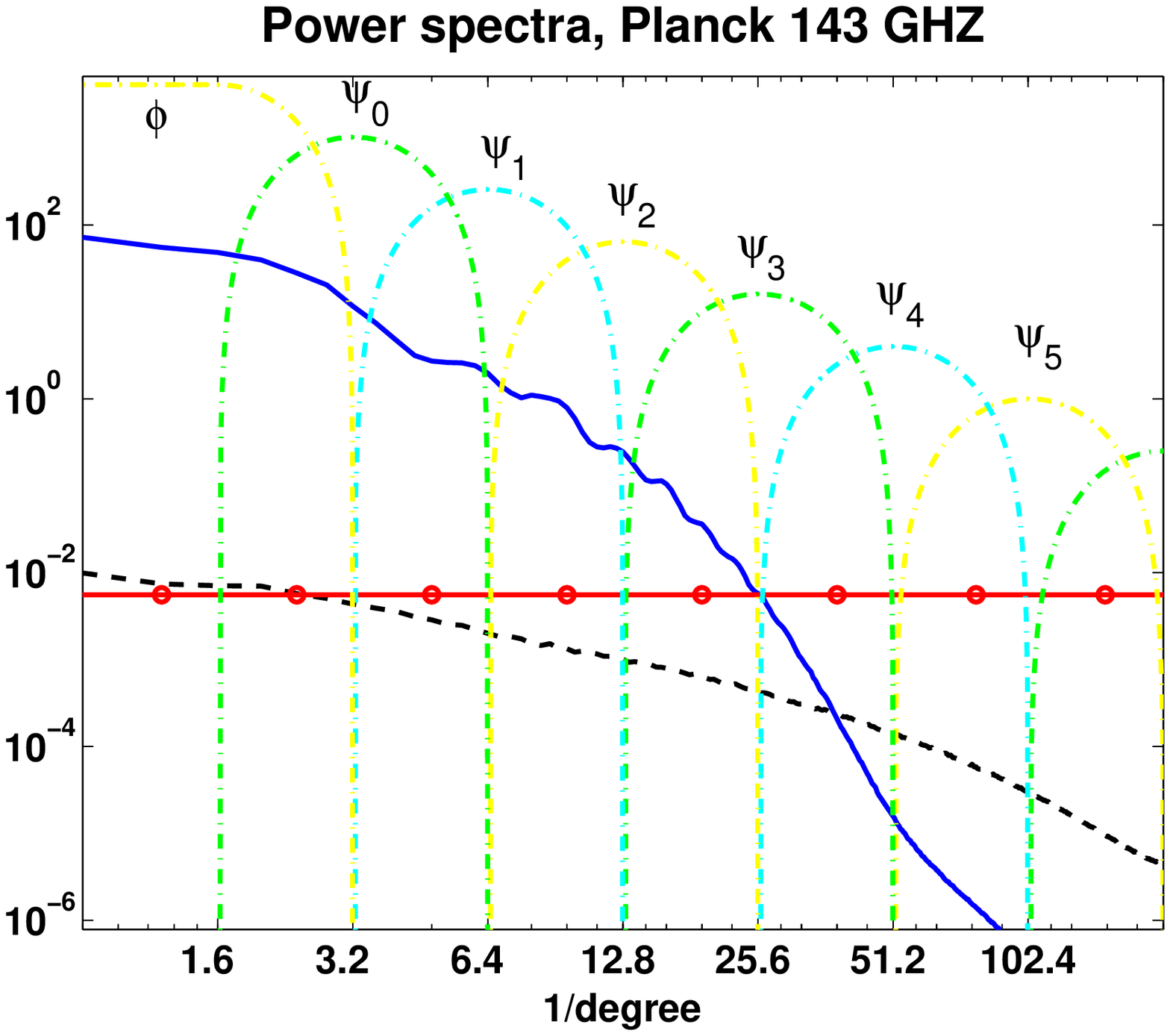}
\includegraphics{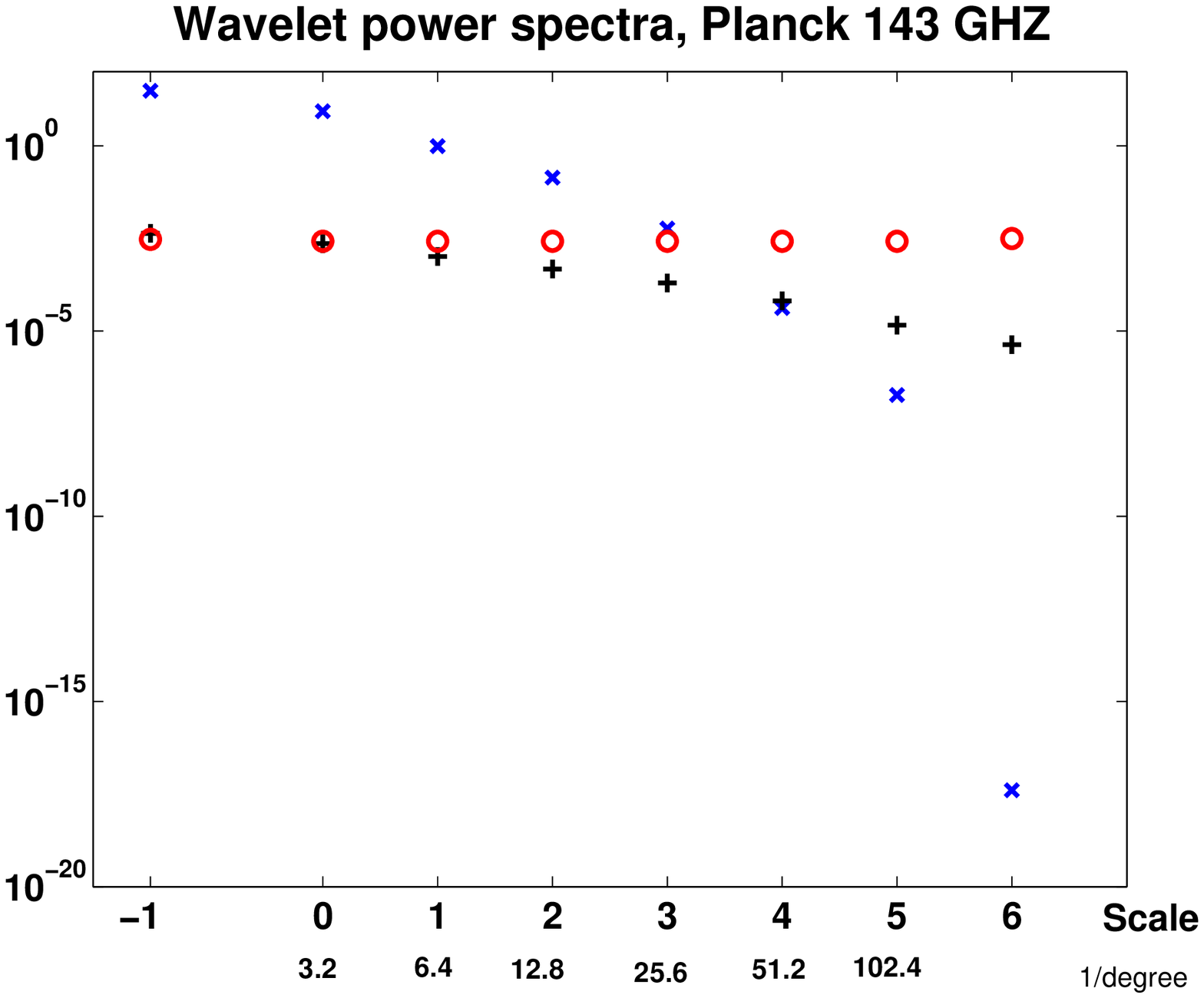}}
\end{center}
\caption{Left: Power spectra in Fourier space  of the wavelets at different scales (dash-dot) and signals : CMB (plain), clusters (dashed) 
and noise (plain with circles) for \Planck\ at 143 GHZ. 
Note: here we only index the scale of the wavelet. 
Right: corresponding Power spectra in wavelet space for the CMB (crosses), clusters 
(plus) and noise (circle) for \Planck\ at 143 GHZ. 
}
\label{fig:power}
\end{figure}

Given an image $s$, it is possible to define a wavelet 
power spectrum $P_w$, which is related to the Fourier based power spectrum 
$P(k) = |\widetilde s (k)|^2$:
\be
P_w(j)=\sum_{m,q} \frac{|\langle s,\psi_{j,m,q}\rangle |^2}{2^{2j}}=
 \sum_{m} \int P(k)\frac{|\widetilde\psi_{j,m,0}(k)|^2}{2^{2j}} dk 
\ee
The Fourier power spectra (in $\mu {\rm K}^2$) for the relevant signals 
(CMB, SZ clusters and noise)
are displayed in Fig.~\ref{fig:power} (left panel)
 together with the windows of wavelets at different scales.
The right panel shows the corresponding wavelet spectra. 

\subsection{The estimator}
Formally, our goal is to estimate several processes (CMB and SZ) from their contributions in observations at different frequencies. We estimate the processes $\{x(p,\nu_0)\}_p$ from the observations $\{y(\nu)\}_\nu$ given that:
\be
y(\nu)= \sum_p{f(p,\nu)x(p,\nu_0)*B(\nu)+N(\nu)}
\label{eq:obs}
\ee
where $x(p,\nu_0)$ is the template of the $p$th process
 at a given frequency $\nu_0$,
$f(p,\nu)$ is the frequency dependence of the $p$th process, 
$B(\nu)$ is the beam (assumed to be a Gaussian with 
given FWHM) and $ N(\nu)$ is the frequency dependent white noise.

In wavelet space, these equations look similar: the observed wavelet coefficients $\langle y(\nu) , \psi_{j,m,q} \rangle = y_{j,m,q}(\nu) $ are linear combinations of the wavelet coefficients of the different processes $x_{j',m',q'}(p,\nu_0)$.

Our estimation method will rely on two principles to discriminate the contributions from different processes. The first one is that we know some statistical properties of the processes (e.g. the CMB and noise are Gaussian processes, while the clusters are not). The second one is that some spatial properties of the processes can be captured by modeling the coherence of wavelet coefficients 
\citep{baraniuk}.
 For example, clusters can be described as spatially localized structures with a high intensity peak. Hence if a cluster is centered at location $q_0$, one should see rather big wavelet coefficients around this location through different scales. If there is no cluster there, all these coefficients should be fairly small. This would not be the case for the noise.
We aim to a local reconstruction which can take advantage 
of these correlations. 
In order to  estimate a particular 
wavelet coefficient $x_{j,m,q}$, we consider a 
neighborhood of coefficients around it: 
\be
{\bf x} \equiv {\bf x}_{j,m,q} =(\ x_{j,m,q}\ ,\ x_{j,m,q+1}\ ,\ x_{j,m,q-1}\ ,\  x_{j-1,m,q}\ ),
\label{neighb}
\ee
such neighborhood contains wavelet coefficients at the same scale 
with close location, and at a close scale with the same location.

Furthermore, we choose to statistically describe the signal as 
a Gaussian scale mixture \citep{Portilla03}:

\be
{\bf x} \equiv  \sqrt z {\bf u}
\label{eq:mod}
\ee
where ${\bf u}$ is a centered Gaussian vector of the same covariance as ${\bf x}$ (${\bf C_x=C_u}$), the multiplier $z$ is a scalar random variable (whose distribution we prescribe later) and the equality holds in distribution. The variables  ${\bf u}$ and $z$ are independent 
and $E\{z\}=1$. 
The covariance of ${\bf x}$ captures the spatial coherence of the process, 
while
the (non-)Gaussianity of the signal is captured by the distribution of the 
multiplier $z$.
We will take the covariance ${\bf C_x}$ to 
be a function of the scale $j$ and orientation $m$.
The convenience of this signal's description will be more clear in the 
next section.

\subsubsection{Step one: de-noising one observation of one process}

To illustrate the idea for the reconstruction process, 
let us first consider the simple case where we observe one process polluted by noise: $y(\nu_0)  =  x(\nu_0) +  N(\nu_0)$. The equations for each single wavelet coefficient and for the neighborhood of wavelets coefficient read:
\bea
y_{j,m,q}  &=&  x_{j,m,q} +  N_{j,m,q}\\
{\bf y}  &=&  {\bf x} +  {\bf N} \nonumber
\eea
where {\bf N}  is Gaussian and {\bf x} is described with the Gaussian mixture
in eq.~\ref{eq:mod}.
The convenience of this representation is that for a fixed multiplier 
  $z=z_0$, the Bayes least square estimate of ${\bf x}$ 
given the observed vector ${\bf y}$ and $z$, would be a Wiener 
filter on the neighborhood of wavelet coefficients: 

\be
E\{{\bf x}|{\bf y}, z=z_0\} = z_0 {\bf C_x} (z_0{\bf C_x} + {\bf C_N})^{-1} {\bf y}.
\ee
However in our model $z$ is not a constant, so $E\{{\bf x}|{\bf y}\}$, the Bayes least square estimate of ${\bf x}$, 
is a weighted average of the Wiener filters above:
\be
E\{{\bf x}|{\bf y}\} = \int_0^\infty{ p(z=z_0|{\bf y}) E\{\bf x} |{\bf y}, z=z_0 \}~dz_0
\label{eq:inte}
\ee
The weights are determined by the probability of $z$ given the observation ${\bf y}$, noted $p(z=z_0|{\bf y})$, which is computed via Bayes rule:
\be
p(z=z_0|{\bf y})=\frac{p({\bf y}|z=z_0) p_z(z_0)}{\int p({\bf y}|z=z') p_z(z') dz'}
\label{eq:pzy}
\ee
where $ p({\bf y}|z=z')$ is a centered Gaussian vector of covariance $z' {\bf C_x} + {\bf C_N}$, and $p_z$ is the probability distribution of $z$ which we will describe in \ref{par:stat}.

Following this procedure, one gets an estimate $E\{{\bf x}|{\bf y}\}$ for each neighborhood of coefficients ${\bf x}$. From this estimated vector, we only keep the estimate of central coefficient $x_{j,m,q}$.

\subsubsection{Step two: de-blurring one observation of one process}\label{par:two}

Consider now the case where the observed signal is a blurred version of the original: $y(\nu_0) = x(\nu_0)*B(\nu_0)+N(\nu_0)$. The convolution with the beam correlates the signal spatially. As a result, a single observed wavelet coefficient is dependent on many wavelet coefficients in the signal. The equations do not decouple any more:
\be
y_{j,m,q} = \sum_{q'} B(\frac{q'-q}{2^j})x_{j,m,q'} +N_{j,m,q}
\label{eq:blur}
\ee

Since support of the beam is infinite, every wavelet coefficient $x_{j,m,q'}$ contributes to $y_{j,m,q}$. However the biggest contribution come from the wavelet coefficients at a close location ($|q-q'|$ small). Hence if the size of the neighborhood is big enough, one can make the following approximation:
\be
{\bf y}_{j,m,q} ={\bf  B_{j} x}_{j,m,q} +{\bf N}_{j,m,q}
\label{eq:approx}
\ee
where $B_{j}$ is a matrix which depends on the beam $B$ and the scale. Eq. (\ref{eq:inte}) and (\ref{eq:pzy}) hold with the modified Wiener filter:
\be
E\{{\bf x}|{\bf y}, z=z_0\} = z_0 {\bf C_x B^{*}_{j}} (z_0{\bf  B_{j}C_x B^{*}_{j}} + {\bf C_N})^{-1} {\bf y}
\ee
and the covariance of $p({\bf y}|z=z')$ is now: $z'{\bf B_{j} C_x B^{*}_{j}} + {\bf C_N}$.

The matrix ${\bf  B_{j}}$ is a truncated version of the matrix 
of convolution by the beam projected at scale $j$.
For eq. (\ref{eq:approx}) to be a good approximation, we allow 
the size of the neighborhood of wavelet coefficients to vary with the scale. 
Specifically, we extend the neighborhood of eq. (\ref{neighb}) 
so that we capture $90 \%$ of the power of the beam at each scale. 
To do this, we simply include in ${\bf x}_{j,m,q}$ the coefficients 
 $x_{j,m,q'}$, with $|q'-q|<k $, choosing $k$ so that 
$\sum_{|q'|<k}| B(\frac{q'}{2^j})|^2>0.9$. Note that the coarser the scale, the smaller $k$ is.

\subsubsection{Step three: de-mixing several processes from several observations}

One can extend this procedure to the case where
several processes contribute to signals observed 
at different frequencies, as in equation (\ref{eq:obs}).

For each process $x(p,\nu_0)$, each neighborhood of wavelet coefficients is modeled as a Gaussian
scale mixture: 
\be
{\bf x}_{j,m,q}(p) \equiv \sqrt{z_{j,m,q}(p)} {\bf u}_{j,m,q}(p)
\ee
where $z(p)$ are scalars of mean 1, ${\bf u}(p)$ are Gaussian vectors, and all these random variables are independent.
The approximation in equation (\ref{eq:approx}) for a neighborhood of wavelet coefficients now reads:
\be
{\bf y}_{j,m,q}(\nu) =\sum_p f(p,\nu){\bf  B_{j}}(\nu) {\bf x}_{j,m,q}(p) +{\bf N}_{j,m,q}(\nu)
\label{eq:approx2}
\ee
Assuming we have $K$ observations and $P$ processes, we note ${\bf Y}=({\bf y}_{j,m,q}(\nu_1),\ldots,{\bf y}_{j,m,q}(\nu_K) )$ and ${\bf Z}=(z(1),\ldots,z(P))$. If ${\bf Z}$ is fixed, we obtain a multicomponent Wiener filter on neighborhood of wavelet coefficients:
\be
E\{{\bf x}(p)|{\bf Y},{\bf Z}\}\!\!=
z(p) {\bf C_x}_{(p)}\!\!\!\!\sum_{k,k'=1}^{K}\!\! f(p,\!\nu_k) {\bf B_{j}^*(\nu_k)} {\bf G}_{k,k'}^{-1} {\bf y}(\nu_k')
\ee
where
\bea
{\bf G}_{k,k'}=\sum_{p=1}^{P} z(p)f(p,\!\nu_k) f(p,\!\nu_{k'}\!) {\bf  B_{j}(\nu_k) C_{x(p)} B^{*}_{j} (\nu_{k'})}\nonumber\\
+\delta_{k=k'}{\bf C_{N(\nu_k)}}
\eea
The Bayes least square estimate of the full model is:
\be
E\{{\bf x}(p)|{\bf Y}\} = \int_{{R}^P_+}{ p({\bf Z}|{\bf Y}) E\{\bf x}(p) |{\bf Y},{\bf Z}\}~dz_1~dz_2..~dz_P
\ee
with the weights:
\be
p({\bf Z}\!=\!(\alpha_i)_i|{\bf Y})=\frac{p({\bf Y}|{\bf Z}\!=\!(\alpha_1,..,\alpha_P)) \prod p_{z_i}(\alpha_i)}{\int\!\! p({\bf Y}|{\bf Z}\!=\!(\beta_1,..,\beta_P))  \prod p_{z_i}(\beta_i)d\beta_i}
\ee
where $p({\bf Y=y}(\nu_k)_k|{\bf Z}\!=\!(\alpha_i)_i)$ is a centered Gaussian with covariance matrix ${\bf G}_{k,k'}$.

In practice we compute the Bayes least square estimate for each wavelet neighborhood ${\bf x}$ and only  keep the central coefficient $x_{j,m,q}$ for each process. We then reconstruct the processes by inverting the wavelet transform.

In order to compute the estimate, we need the covariance matrices ${\bf C_{N(\nu)}}$ and ${\bf C_{x(p)}}$. 
Note that these matrices depend on the scale and orientation .
Since the level of noise is assumed to be known, the covariances 
${\bf C_N(\nu)}$ can be computed. The covariances for the different 
processes ${\bf C_{x(p)}}$ are estimated from simulated input maps.
 We use the wavelet transform described in (\ref{par:wavdec}) with 4 
orientations and 5 scales for the ACT experiment, with 4 orientations 
and 6 scales for the \Planck\ experiment. The size of the neighborhoods 
is chosen {\it adaptively} at each scale so that the approximation in
 equation (\ref{eq:approx2}) is valid (as explained in \ref{par:two}). Finally, we choose the probability 
distributions $p_{z_i}$ to capture the properties of the process $x(i,\nu_0)$, 
as noted in the next section.

\subsection{Statistical properties of the signals}\label{par:stat}

We now must decide which distribution $p_z$ we intend to use for 
each signal.
In the case where $z\equiv 1$, the Gaussian scale mixture described in eq. (\ref{eq:mod}) reduces to a Gaussian process. 
We are considering here the CMB to be Gaussian,
and will consistently assume $z\equiv 1$ at all scales for this signal.
 
The cluster signal however is typically non-Gaussian. 
In order to model such non-Gaussianity, we will need a 
more elaborate distribution for $z$, which could in principle 
be chosen with or without any particular link to the true distribution.
In the following we will consider different cases for the cluster's $z$
distribution (with and without a physical significance) and we will compare
the results.

In order to compute a physically based model for the cluster non-Gaussianity,
we analyzed the simulated SZ map.
For the Gaussian scale mixture model of the signal,
 it is difficult to solve the closed form equation for $p_z$, 
(the probability of $z$) given the probabilities of $x$ and $u$.
 We will instead study the probability of its logarithm: 
 $p_{\ln z}$ and refer to it as the prior. 
Note that $p_z$  can then be easily recovered:
 $p_z( v ) dv  = p_{\ln z} ( u ) e^u du $ for $u=\ln v$.
Taking the the logarithm of eq. (\ref{eq:mod}), one obtains the following equation:
\be
p_{\ln |x|}(v)=\int_0^\infty 2~p_{\ln z}(2y)~p_{\ln |u|}(y-v)~dy
\label{eq:plnz}
\ee
which we solve given that $u$ is Gaussian and $p_{\ln |x|}$ is estimated from the SZ input maps (details will be given in an upcoming paper (Anthoine, .
 \textit{in prep.}).\nocite{Anthoine_in_prep}

Together with the physical prior described above,
 we will also consider two other non-physical ones. These are:
{\it i)} the Gaussian prior, which
 corresponds to assuming $z\equiv 1$, or equivalently $p_{\ln z}(v)=\delta_D(v-0)$;  {\it ii)} the non-informative prior, i.e.
the uniform probability distribution on $\ln z$, $p_{\ln z}(v)= \textit{constant}$. 
The latter  has been used for recovering non-Gaussian 
signals in digital image processing.
We will compare the performances of the non informative  and Gaussian 
prior to the one  that we derived from the input maps.

\begin{figure}
\begin{center}
\includegraphics[width=\columnwidth]{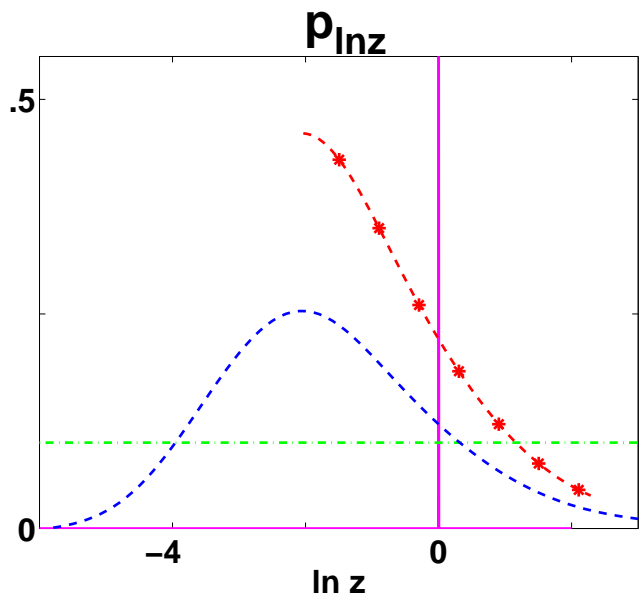}
\end{center}
\caption{Examples of probability distribution for $\ln z$ for a particular 
scale of the wavelet decomposition.
 Here $z$ is the multiplier in the Gaussian scale mixture model. 
Plain: Gaussian prior, $p_{\ln z}(v)=\delta_D(v-0)$, $z \equiv 1$. 
Dash-dot: non-informative prior, $p_{\ln z} \equiv constant$. 
Dashed: {\it profile} computed from the data. 
Dashed and stars: profile computed from the data truncated ({\it profile d}).}
\label{fig:pz}
\end{figure}

We plot the distributions described above in  Fig.~\ref{fig:pz} for the scale
where the SZ signal is most pronounced.
The behaviour of $p_{\ln z}$ for small (resp. large) values influences the probability of finding small (resp. large) values of $|x|$.
 In our model, $p_x$ describes the statistics of wavelet coefficients. Hence, the higher the probability 
$p_x$ for small values of $|x|$, the sparser we expect the signal to be. The 
more pronounced the tails of $p_x$ for large 
$|x|$ are, the more intense we expect the non-zero signal (i.e. the clusters) to be.

\begin{figure}
\begin{center}
\includegraphics[width=\columnwidth]{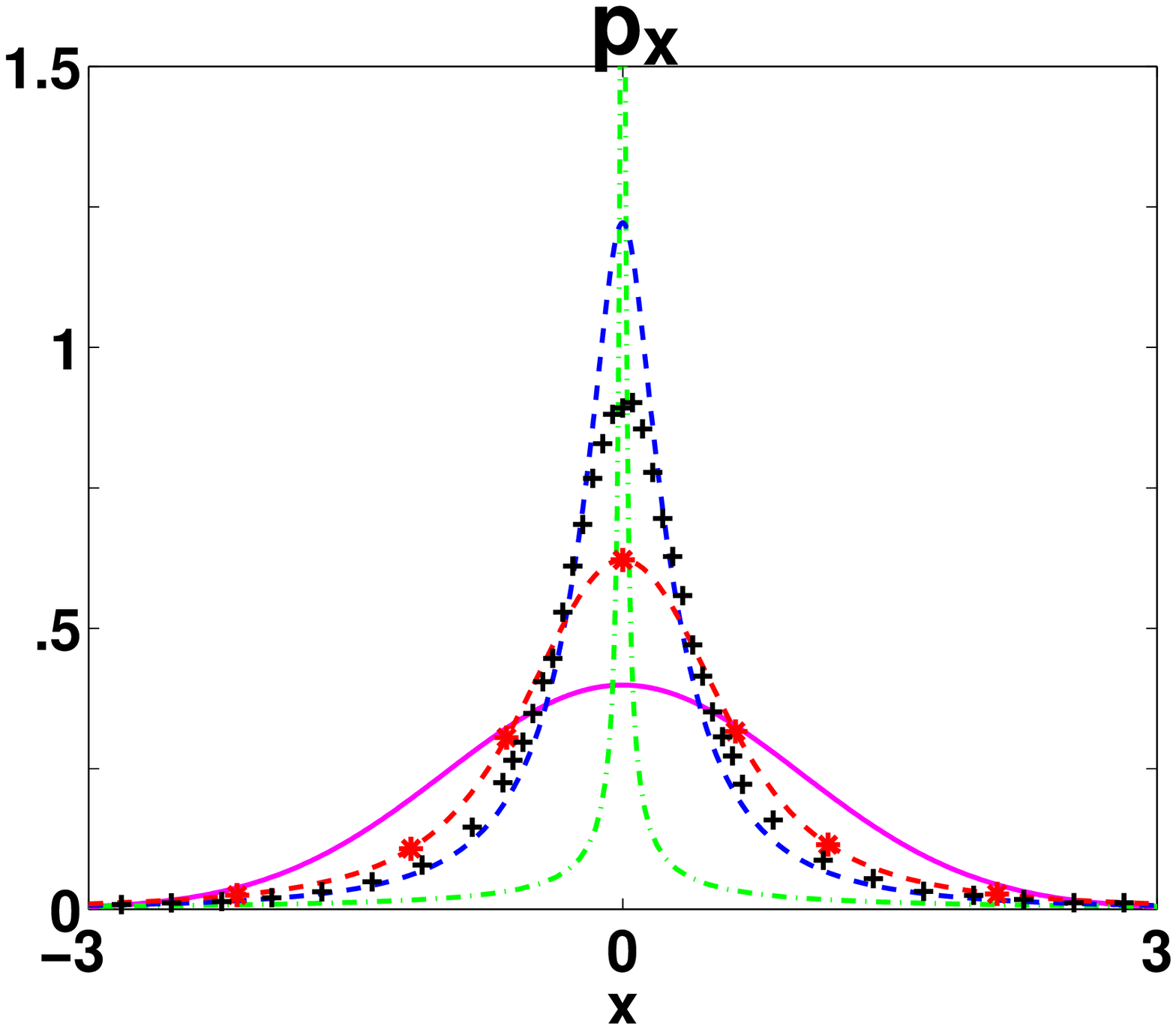}
\includegraphics[width=\columnwidth]{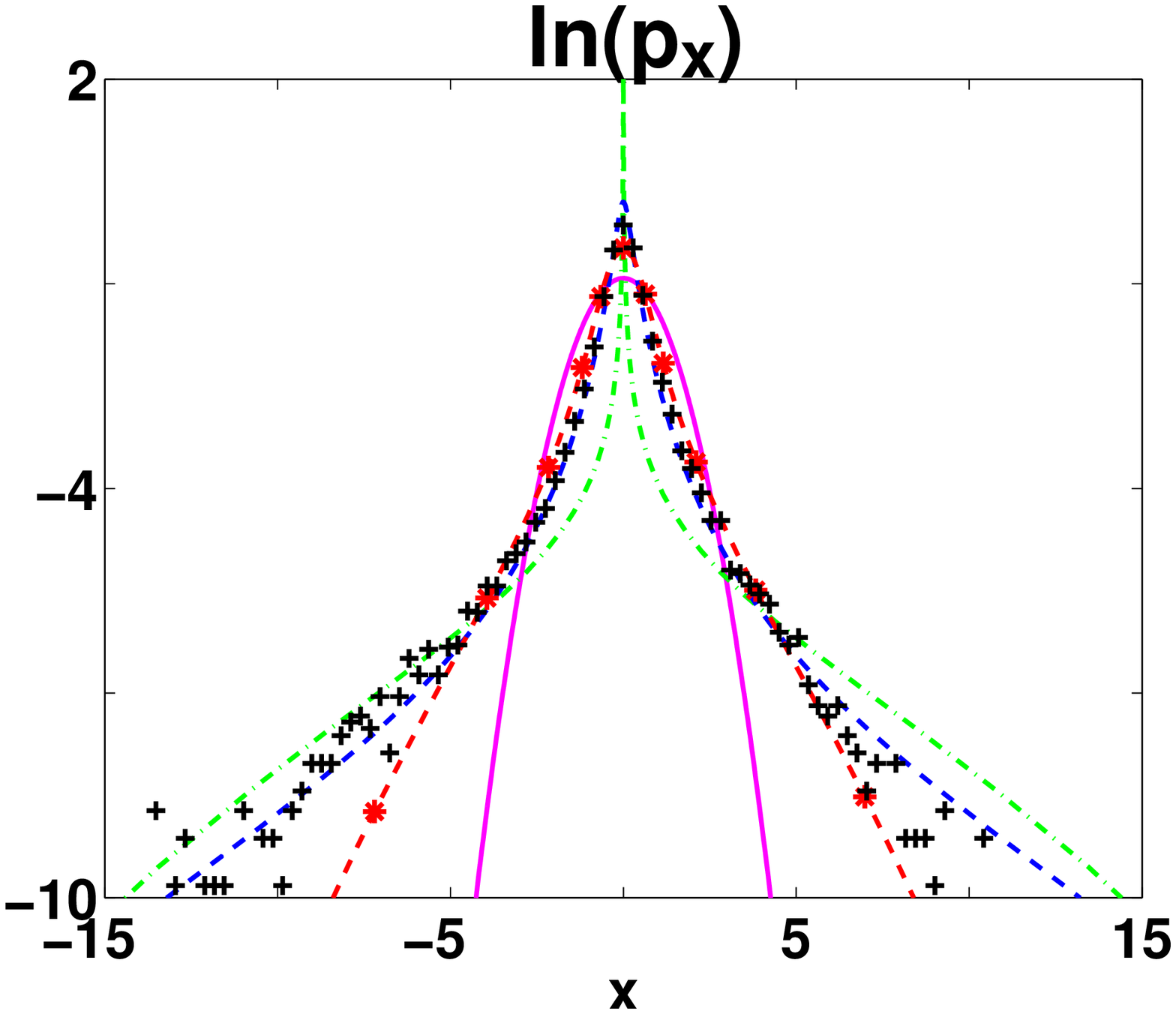}
\end{center}
\caption{Top panel: $p_x$, the distribution of $x \equiv \sqrt{z} u$.
 Bottom panel: the logarithm of this distribution: $\ln(p_x)$. 
In order of increasing value of $p_x$ at $x=0$ (left) and increasing value of $|x|$ for $\ln(p_x)= -10$ (right).
Plain: Gaussian prior, x is Gaussian.
Dashed and stars: $p_x$ corresponding to the profile computed from the data truncated ({\it profile d}).
Plus: distribution $p_x$ as numerically computed from the input maps.
Dashed: $p_x$ corresponding to the {\it profile} computed from the data. 
Dash-dot: $p_x$ corresponding to the non-informative prior.}

\label{fig:xdist}
\end{figure}

In Fig.~\ref{fig:xdist}, we plot the marginal distribution of the wavelet coefficients for SZ clusters as computed from the map, and show how the different 
$z$ priors  in fig.~\ref{fig:pz} would fit the actual distribution. 
 The right panel shows that the tails of the clusters' distribution is much 
broader 
than the one of the Gaussian. As a consequence, the Gaussian prior tends 
to underestimate the amplitude of clusters. The left panel shows that the 
Gaussian prior underestimates the number of wavelets coefficients with 
small amplitude, while the non informative prior and to some 
extent the profile prior overestimate them. This means that the sparsity 
of the cluster signal is not well described by these distributions. 
The Gaussian prior will tend to create more clusters than necessary 
while the non informative prior or the profile will miss clusters. 

We will also discuss  another case that is derived from the physical
one:  
 a truncated version of the  $p_{\ln z}$ estimated from the input maps  
(which we refer to as {\it profile d}). This version is a trade-off 
between the profile computed from the data and the Gaussian prior.

To summarize:
in the rest of the paper we will compare the results given by these four priors:\\
\indent a) Gaussian prior, $p_{\ln z}(v)=\delta_D(v-0)$, referred to as {\it Gaussian}.\\
\indent b) Non informative prior, $p_{\ln z}(v)= constant$, referred to as {\it Uniform}.\\
\indent c) Profile computed from the data, referred to as {\it Profile}.\\
\indent d) Truncated profile computed from the data, referred to as {\it Profile d}.

\section{Results} \label{par:res}

\subsection{The reconstructed maps   }

\begin{figure}~
\begin{center}
\includegraphics[height=0.2\textheight]{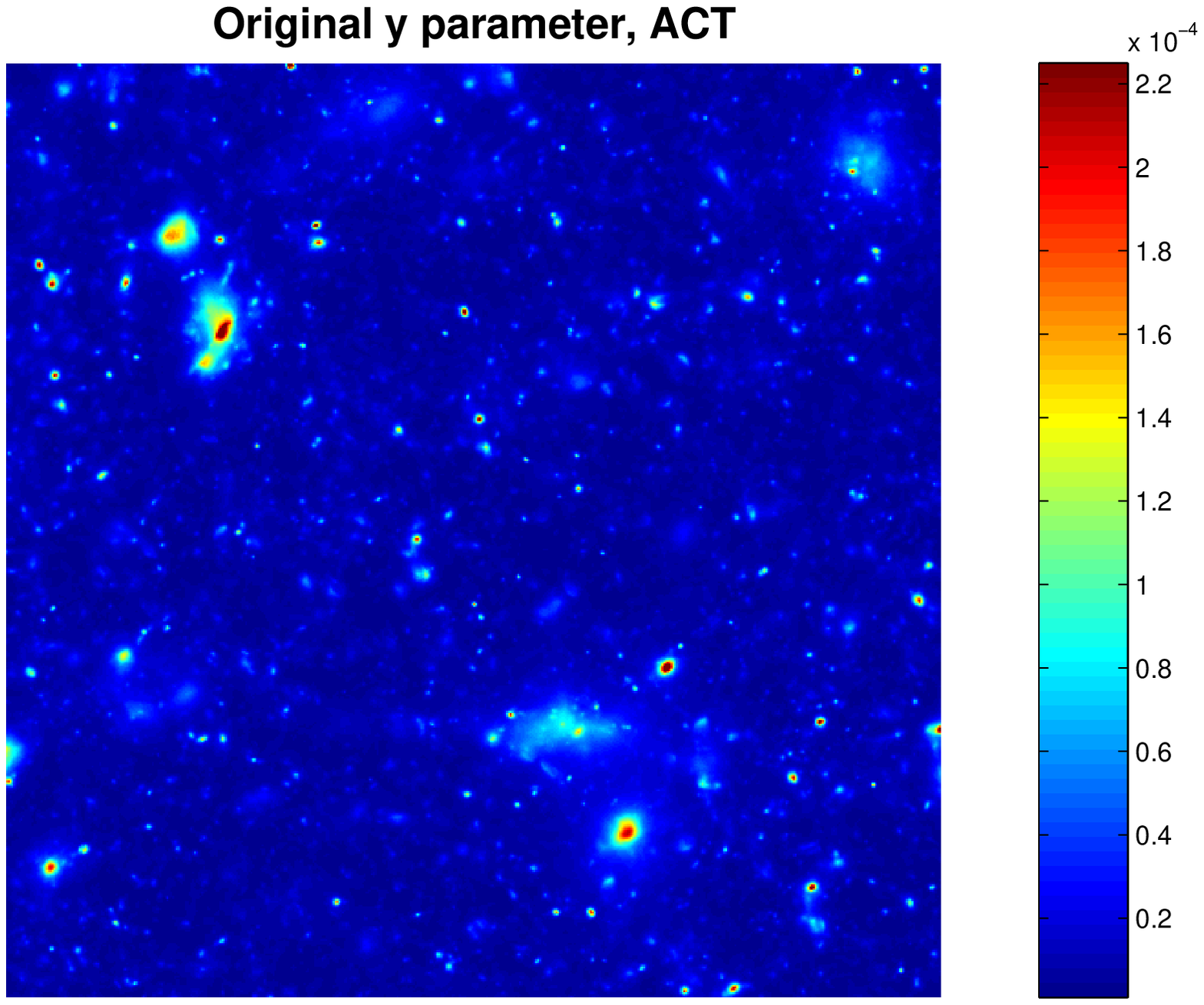}
\includegraphics[height=0.2\textheight]{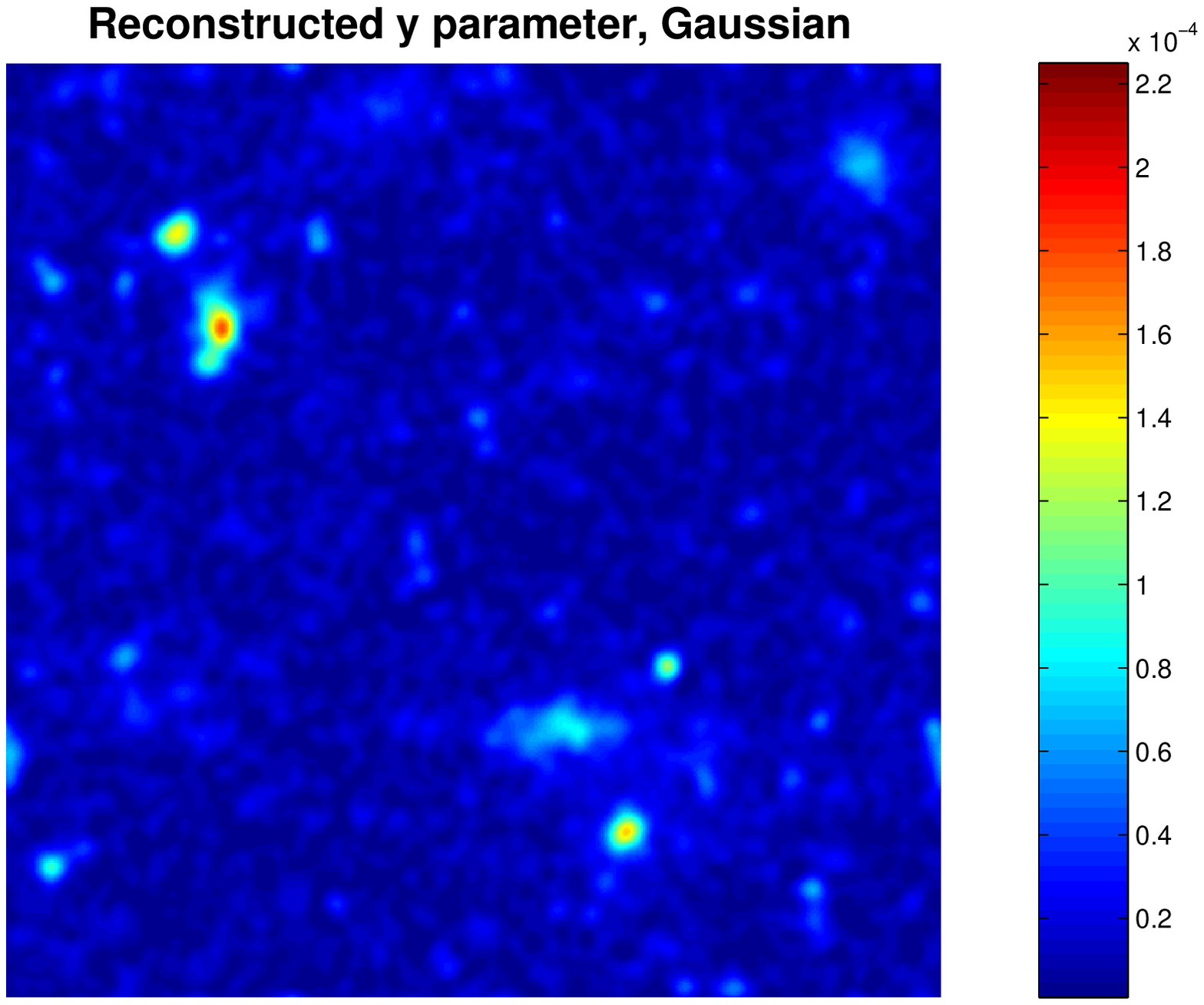}
\includegraphics[height=0.2\textheight]{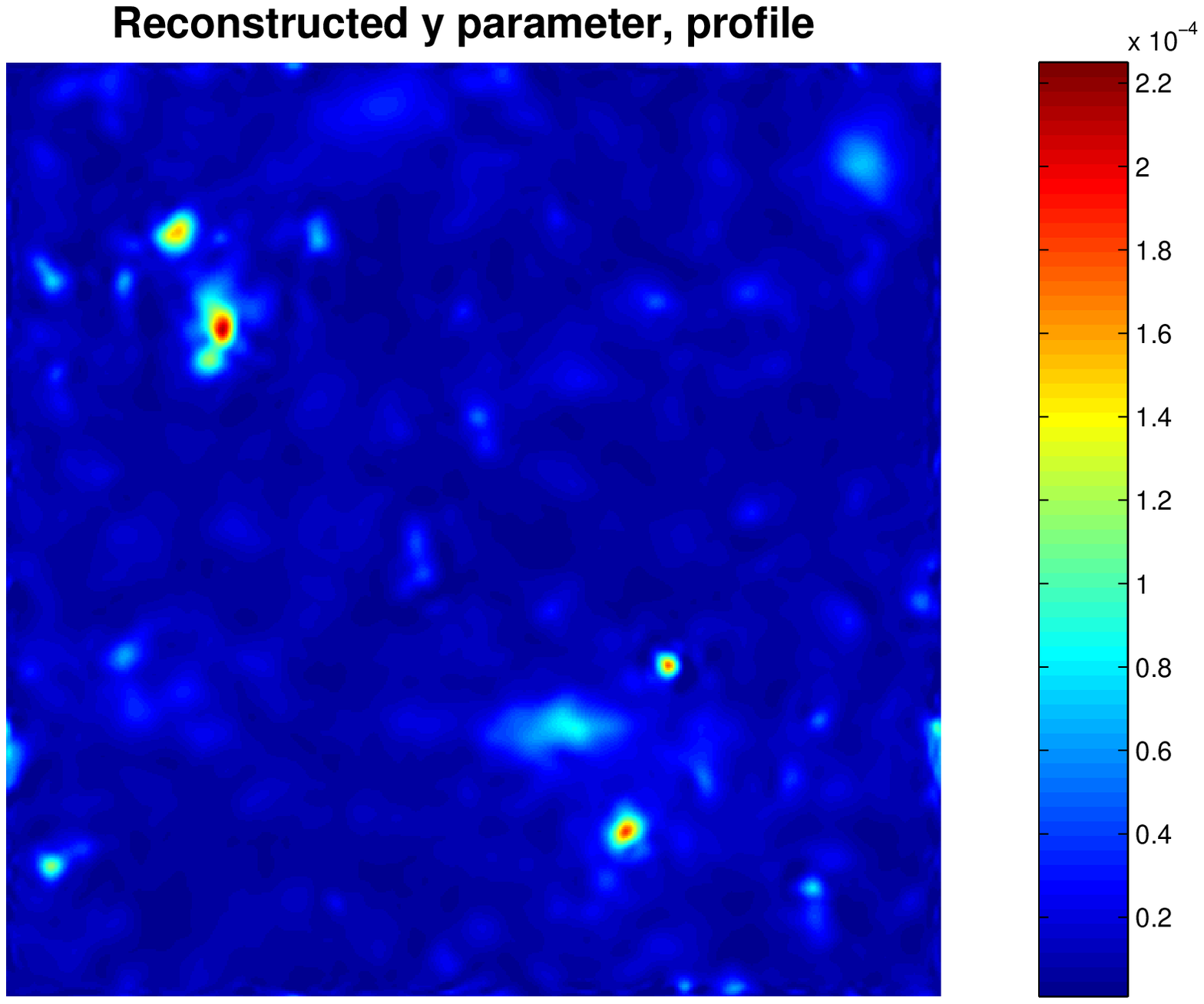}
\includegraphics[height=0.2\textheight]{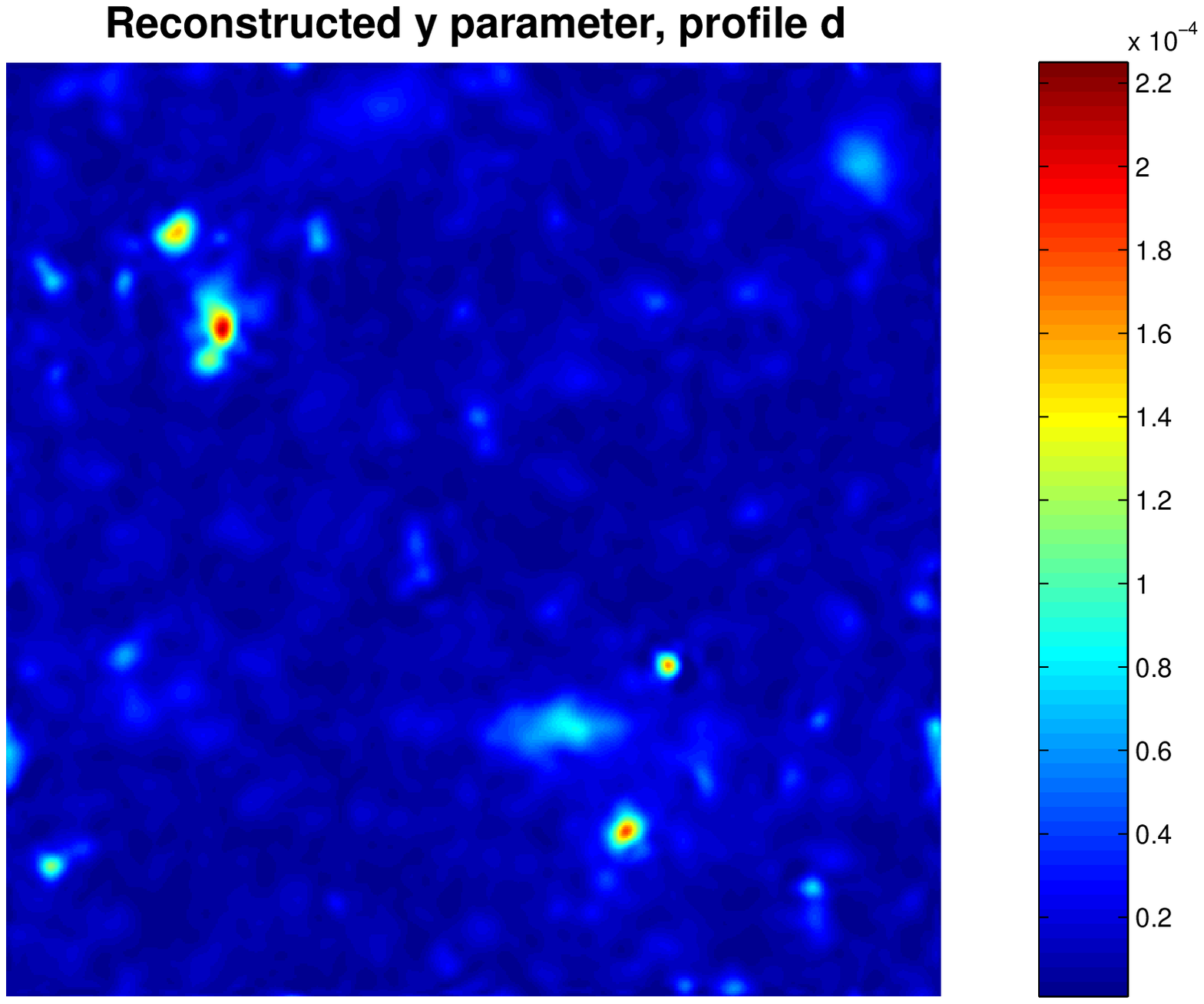}
\end{center}
\caption{ The simulated (top panel) and reconstructed ACT cluster maps. 
The second panel correspond to the Gaussian prior, the third to the
actual cluster prior {\it profile}, and the fourth to the truncated cluster prior {\it profile d}.  The maps is 1.2 deg on a side.}
\label{fig:mapsACT}
\end{figure}

\begin{figure}~
\begin{center}
\includegraphics[height=0.2\textheight]{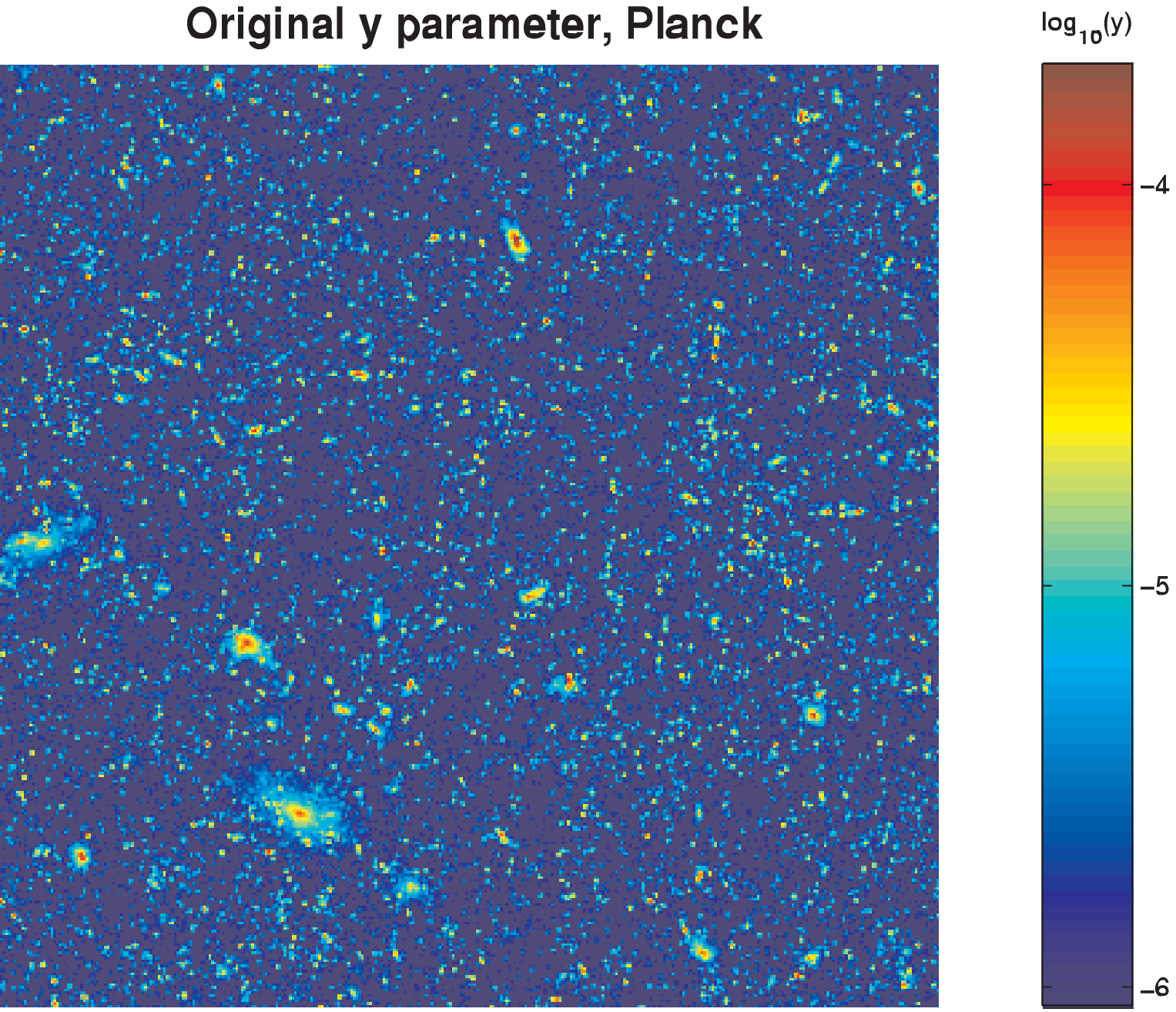}
\includegraphics[height=0.2\textheight]{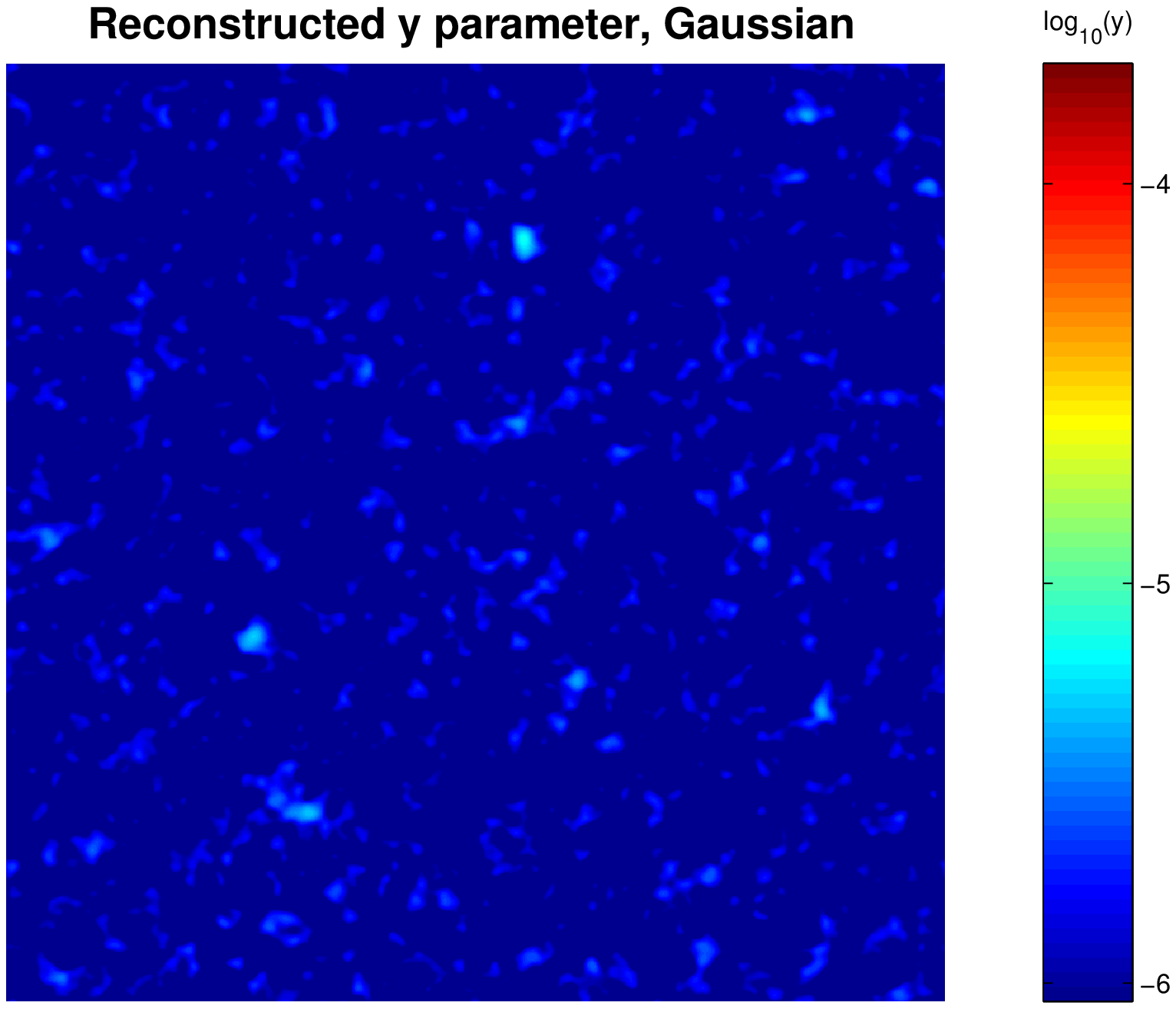}
\includegraphics[height=0.2\textheight]{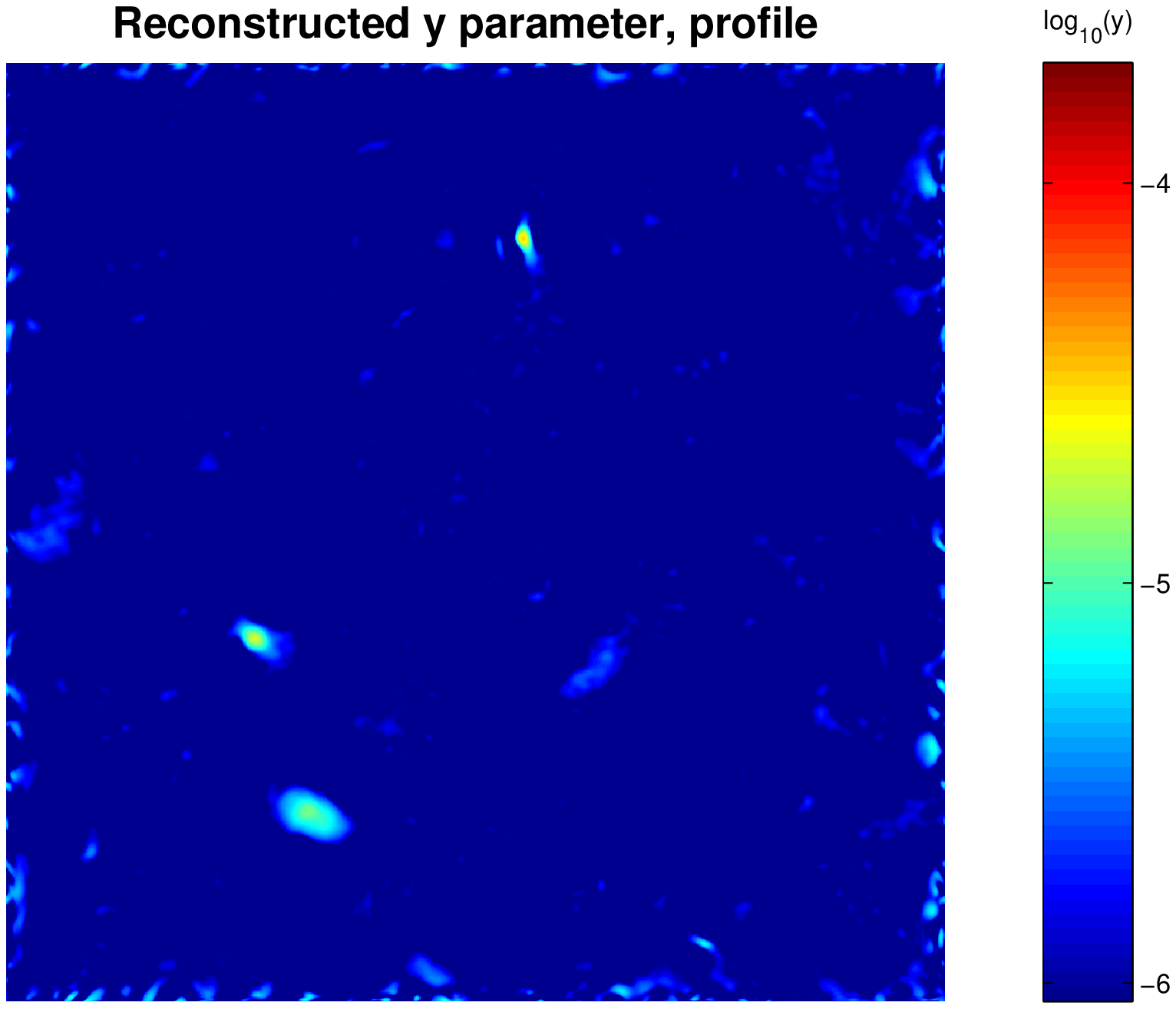}
\includegraphics[height=0.2\textheight]{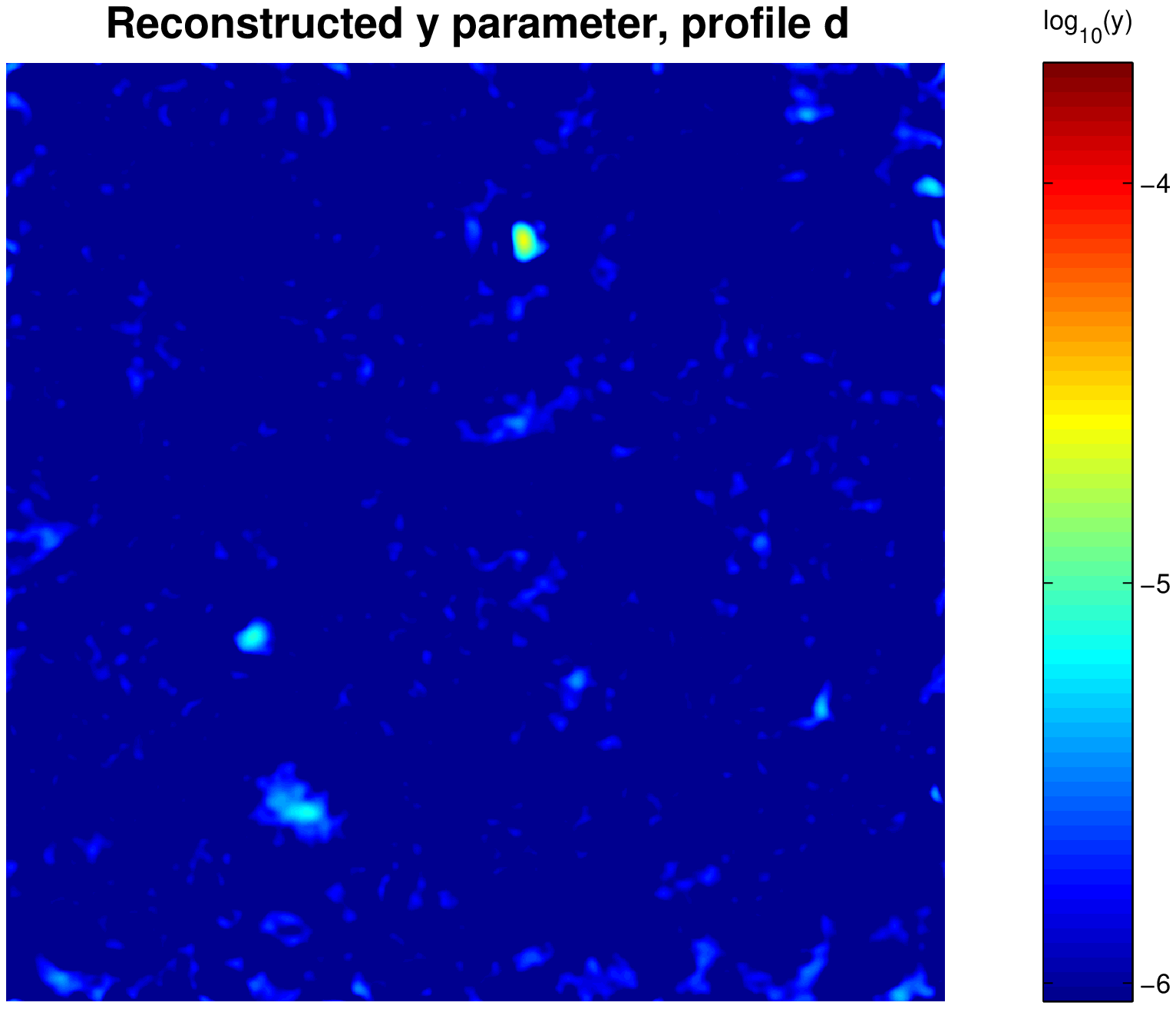}
\end{center}
\caption{ The original and reconstructed \Planck\ cluster maps. 
The distributions are the same as corresponding ACT panels.
Note that the 
color scale here is in the logarithm of the luminosity.
Border effects are visible: the areas at the edges 
are disregarded in the analysis.  The map is 10 deg on a side.}
\label{fig:mapsPlanck}
\end{figure}

We applied the method described above to the ACT and \Planck\ simulated maps
with realistic beam and noise properties (see table~\ref{tab:ic}),
 with different assumptions on 
the prior $p_{\ln z}$. In figure~\ref{fig:mapsACT} and~\ref{fig:mapsPlanck}
we show the input and reconstructed $y$ maps for ACT and \Planck\ respectively.

In these figures we notice that, even using a wavelet basis which allows for
a local reconstruction, the Gaussian prior for $p_{\ln z}$ 
(which causes the estimator to reduce to Wiener filtering on the 
neighborhood of wavelet coefficients) underestimates the
 high-intensity peaks.
On the contrary,
the profile and uniform prior perform (equally) better 
reconstruct the central intensity of the bright clusters,
indicating that the specific shape of the tail of the prior 
distribution of multipliers for high $z$
is not particularly relevant, as long as it provides
 enough power at sufficiently high $z$ values.

As for the low-intensity clusters, they seem better reconstructed
in the Wiener filtered maps than by using the real profile (or uniform) prior
for $p_{\ln z}$.
This indicates that our estimate of the true $p_{\ln z}$, 
the {\it profile}, may be weighing too heavily  
low-intensity points.
This effect could be caused by the procedure we adopt in 
computing $p_{\ln z}$  from the input maps.
Because of our deconvolution technique (see section \ref{par:stat}, 
eq.~\ref{eq:plnz}),
it may well be that if the true $p_{\ln z}$ had a very sharp drop at some low-$z$
value, we wouldn't be able to model it accurately.
For this reason, we also tried a profile that corresponds to the true one 
truncated at $z$'s lower than the peak point (see fig.~\ref{fig:pz}, {\it profile d} case).
The truncated profile  performs as well as Wiener filtering in recovering 
low-intensity clusters, still improving the results on high-intensity ones.

\subsection{The central $y$ parameter}

\begin{figure}
\begin{center}
\includegraphics[width=\columnwidth]{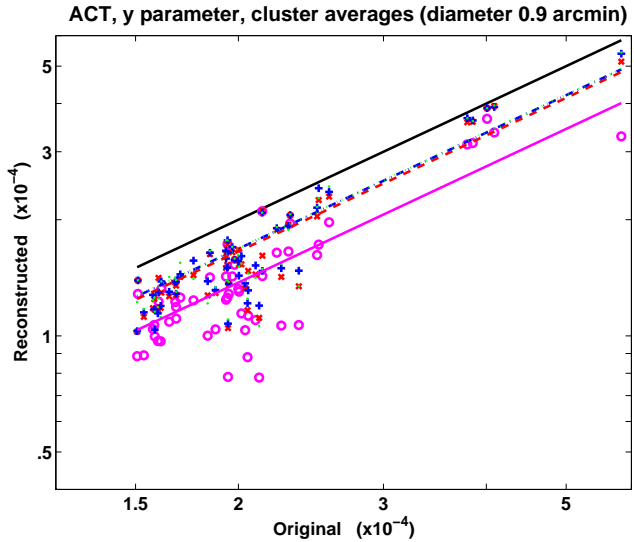}
\end{center}
\caption{ The input and output $y$ parameters for bright clusters in the ACT
experiment. Top plain line: line of perfect reconstruction. Gaussian prior: circle, bottom plain line; Profile prior: plus, dash-dot line; Profile d prior: cross, dashed line.} 

\label{fig:slopeACT}
\end{figure}

\begin{figure}
\begin{center}
\includegraphics[width=\columnwidth]{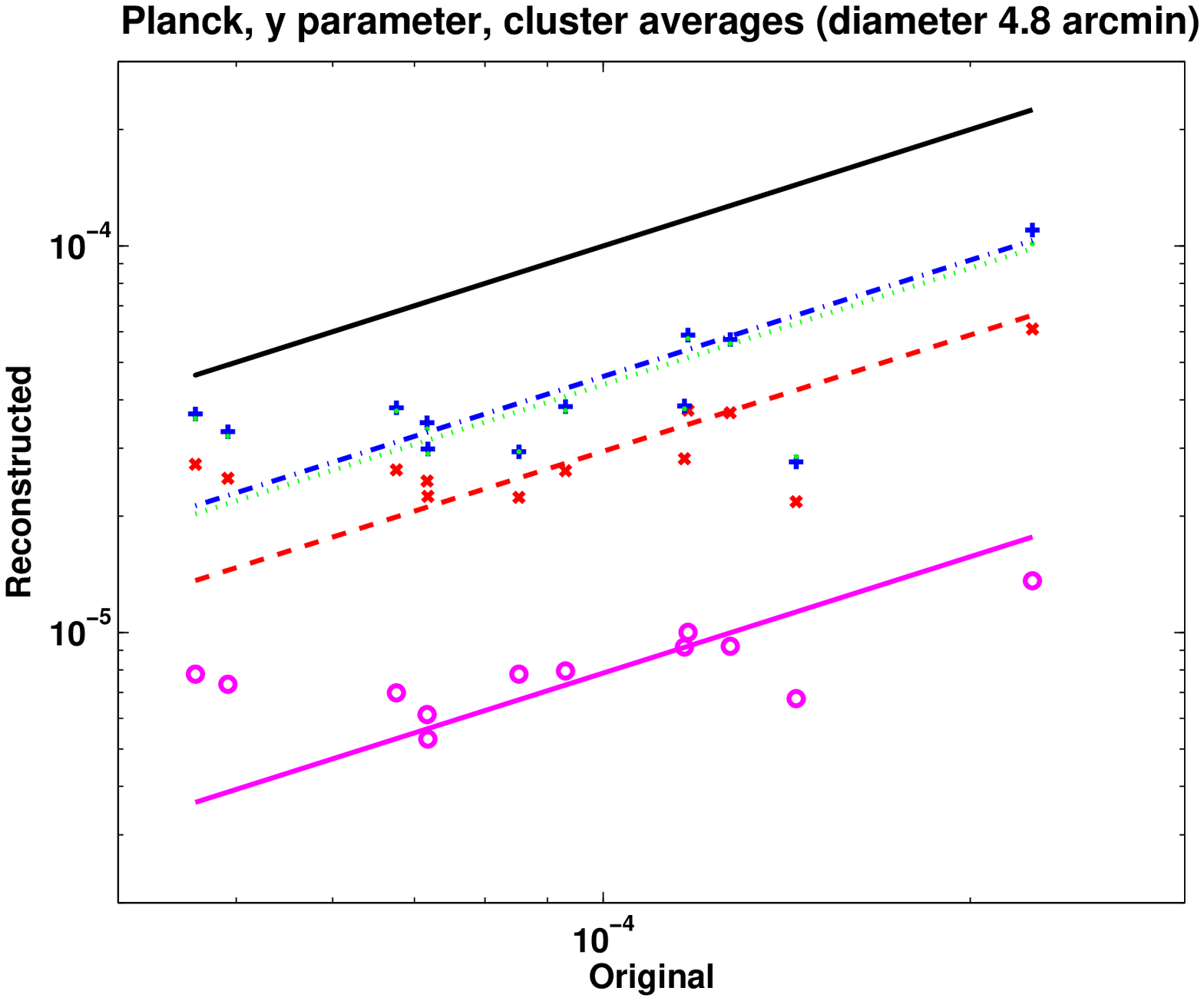}
\end{center}
\caption{ The input and output $y$ parameters for bright clusters in the 
\Planck\ experiment.  Top plain line: line of perfect reconstruction. Gaussian prior: circle, bottom plain line; Profile prior: plus, dash-dot line; Profile d prior: cross, dashed line.}

\label{fig:slopePlanck}
\end{figure}

When it comes to infer cosmological parameters from number counts in other 
wavebands (e.g. X-ray and optical)
the common practice is to  
retain only the brightest clusters which are less affected
by selection effects and have a better characterized scaling function.
We shall adopt the same strategy with SZ clusters, also motivated by the fact that they are less affected by reconstruction errors.
For these clusters it is appropriate to assess how the input $y$ parameter
relates to the reconstructed one.
In figs.~\ref{fig:slopeACT} and \ref{fig:slopePlanck}
  we show the reconstructed $y$ parameter versus the input one for ACT and 
\Planck\ respectively, smoothed over a scale which is roughly the smallest 
beam size in each experiment.
In table \ref{tab:slsp}, we quote the slope of the fitting line for the different distributions considered,
 together with the average percentage departure from it, or spread, 
computed as the mean of the ratio: $|(\textrm{input} - \textrm{predicted})/\textrm{input}|$ with 
$\textrm{predicted} = \textrm{output} / \textrm{slope}$.
In general, using either the uniform or the {\it profile} prior for 
$p_{\ln z}$ improves
both the average reconstruction (the slope of the curve) and the associated error. The actual performance depends on the intensity cut applied:
here we plot the 50  brightest ACT clusters and the 12 \Planck\ most 
extended clusters in the reconstructed  maps (see section~\ref{sec:pco}
for a discussion on selection effects).
In the case of ACT the average reconstruction improves by about 20 per cent 
on the Gaussian prior case when our approach is applied using any of the three other distributions.
The scatter is also reduced by a factor 2.
In the case of \Planck, the improvement in the average reconstruction is roughly a factor 5, and the error is also reduced.

\begin{table}
\begin{center}
\begin{tabular}{cccccc}
& & Gaussian & Uniform & Profile & Profile d\\
\hline 
\hline
Slope & ACT & 0.69 &   0.84 &   0.84 &   0.83 \\
&\Planck\ &  0.07 &   0.41 & 0.43 &   0.27  \\
\hline
\hline
\hspace{-1mm}Spread & ACT &   0.15 &   0.12  &  0.12  &  0.11 \\
&\Planck\ & 0.35 & 0.27 & 0.27 & 0.31
\end{tabular}
\end{center}
\caption{ The slope and the spread of the the lines in
figs~\ref{fig:slopeACT} and \ref{fig:slopePlanck}, obtained
with different priors $p_{\ln z}$.
}
\label{tab:slsp}
\end{table}

\begin{figure}
\begin{center}
\includegraphics[width=\columnwidth]{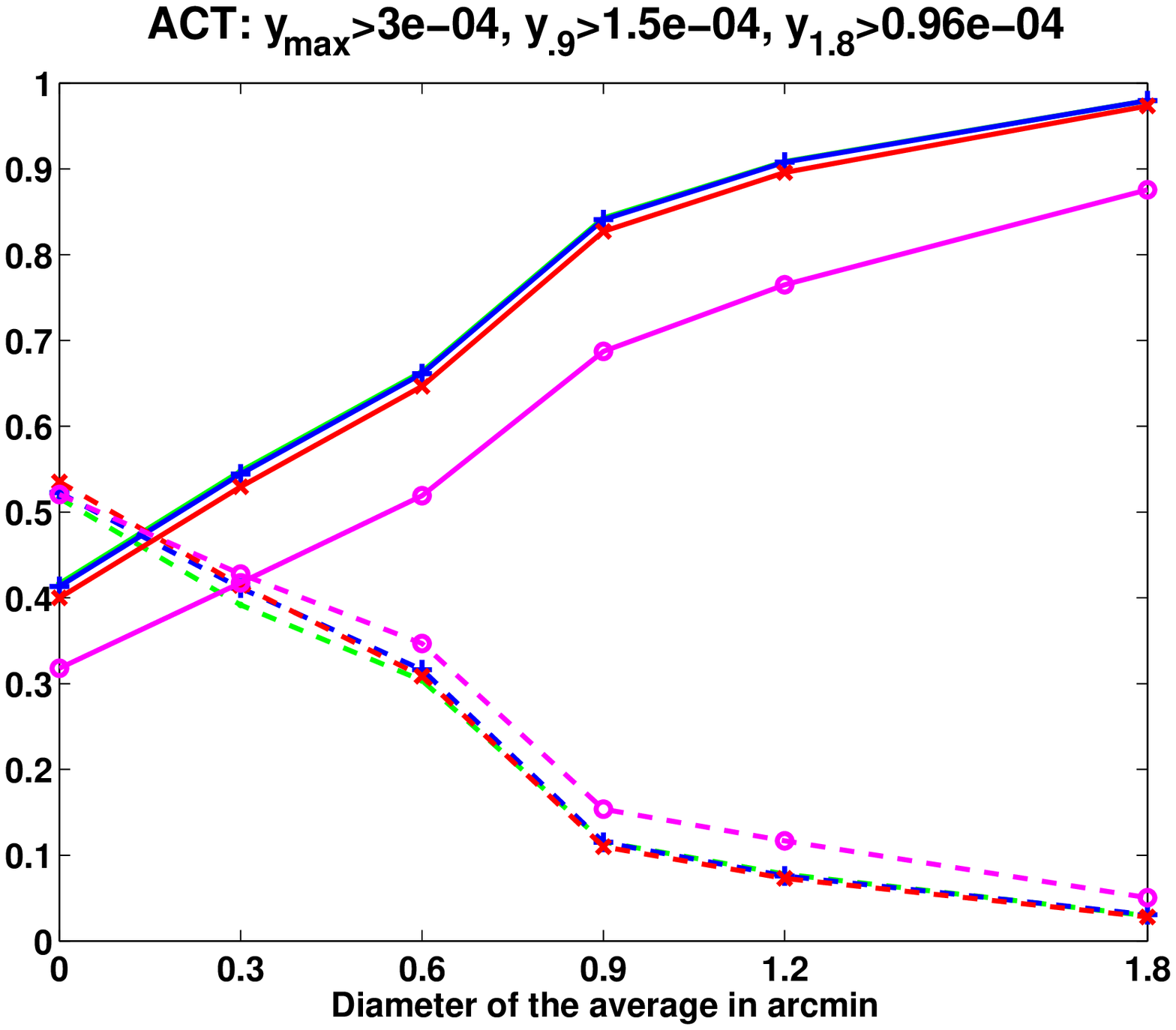}
\end{center}
\caption{ The slope (plain) and spread (dashed) of the fitting line for the input/output
$y$ parameter when averaged over different angles $\theta_c$.
This plot is constructed using the 50 brightest clusters in 24 $(1.2~{\rm deg})^2 $ ACT  maps.}

\label{fig:slopespreadACT}
\end{figure}

\begin{figure}
\begin{center}
\includegraphics[width=\columnwidth]{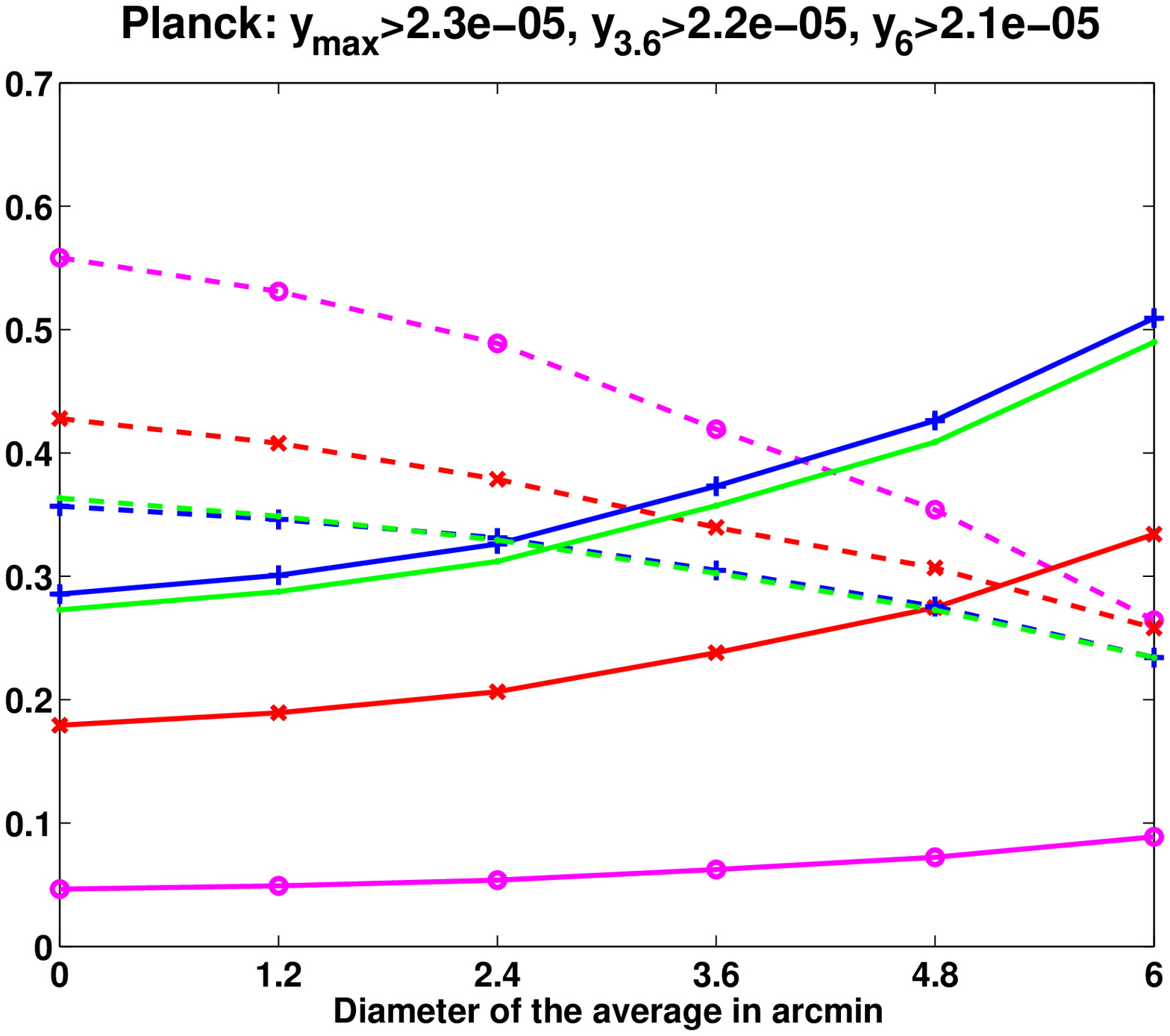}
\end{center}
\caption{ The slope (plain) and spread (dashed) 
of the fitting line for the input/output
$y$ parameter when averaged over different angles $\theta_c$.
This plot is constructed using the 12 biggest clusters in $(10~\textrm{deg})^2$  \Planck\  maps.}

\label{fig:slopespreadPlancktrue}
\end{figure}

\begin{figure}
\begin{center}
\includegraphics[width=\columnwidth]{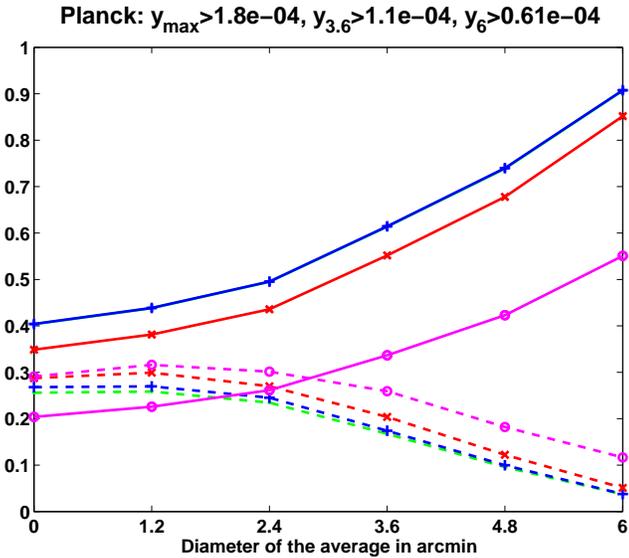}
\end{center}
\caption{ The slope (plain) and spread (dashed)
 of the fitting line for the input/output
$y$ parameter when averaged over different angles $\theta$. 
This plot is constructed using the 25 brightest clusters in the 10 \Planck\ 
  maps. In this case, the noise has been reduced by a factor 7 in all frequency bands.}

\label{fig:slopespreadPlancknonoise}
\end{figure}

The profile and uniform prior perform very similarly in reconstructing the
cluster centers, so that we only quote the reconstruction parameters 
for the profile prior. It is comforting, however, that the details in the
shape of the non-Gaussian profile play a minor role in the
 reconstruction performance.

\subsection{Performances in the $y$ parameter reconstruction}

We now come to the  relevant question: which observable should be used in 
order to derive cosmological parameters? In particular which should be the 
angle $\theta_c$ over which the $y$ parameter should be averaged?
In general we expect the answer to depend on the specific 
 experiment in hand.
In order to answer this question, we smoothed the input and reconstructed
maps over several angles, and then we computed the slope of the reconstruction
and the spread of the points around that slope.

In figures ~\ref{fig:slopespreadACT} and \ref{fig:slopespreadPlancktrue}
 we show the slope and spread of the 
fitting as a function of the smearing angle for the ACT and \Planck\ experiments, when the clusters in figs.~\ref{fig:slopeACT} and \ref{fig:slopePlanck} are considered.

As for the ACT case, we notice  a big improvement when 
we reach the highest resolution of the instrument. At this $\theta_c$
both the slope increases significantly (yielding almost perfect 
reconstruction: 0.8--0.9)  and the spread is greatly reduced.
Notice that for ACT the method proposed here reduces the spread by about a 
factor 2. The overall error in the Reconstruction (10 per cent)
should not present a major impediment in deriving cosmological parameters
from this sample.

\begin{figure}
\begin{center}
\includegraphics[width=\columnwidth]{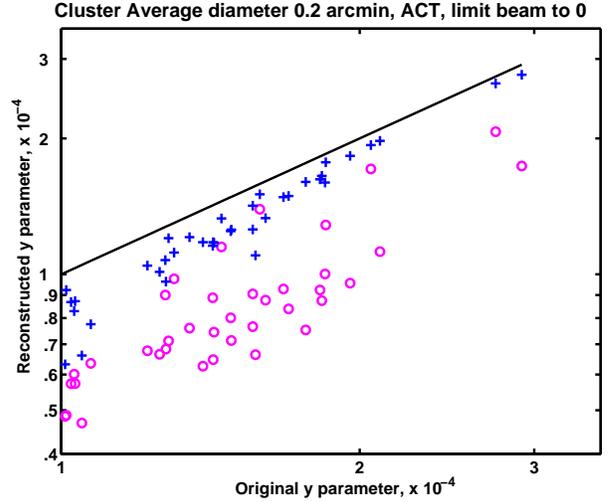}
\end{center}
\caption{ Reconstruction of the $y$ parameter in the ACT experiment 
in the idealistic case where  the beam is set to zero. The {\it profile} prior (plus) 
significantly reduces the spread in the reconstructed $y$ parameter with 
respect to the Gaussian prior (circle).}

\label{fig:beam0}
\end{figure}

The beam size is also driving the spread associated with the slope, which drops
dramatically for $\theta_c \simeq 0.9$.
In order to better understand the role of the beam, we studied the idealistic 
case in which the beam is virtually zero (see fig.~\ref{fig:beam0}).
We find that in this limit, the Wiener filter reconstruction
 still shows a spread, while in our proposed model the spread is minimal. 
We conclude that should observational performances improve in the upcoming 
years, this method would become even more  interesting.

Also in the case of  \Planck\ (fig.~\ref{fig:slopespreadPlancktrue}), 
 we  see a 
significant 
improvement of this method on Wiener filtering in terms of the slope and the 
spread. However, the spread is still about 30 per cent even for the most 
extended  clusters considered here.
Note also that these clusters were selected by being the biggest in the sample
rather than the brightest, because some of the very compact
bright clusters get completely confused with noise and do not get reconstructed
at all.
 This could be a potential problem when it comes to derive cosmological
parameter from them. 
However, the real performances of the \Planck\ instrument may be better than the
ones used in this simulation. Moreover, the noise in the sky will not 
be uniformly distributed, in fact some areas will be better sampled than 
others. We assessed the relevance of the noise level by 
performing a similar analysis on the \Planck\ maps with a reduced
level of noise (about a factor 7 lower).
The results are displayed in fig.~\ref{fig:slopespreadPlancknonoise} for the 
25 brightest clusters in the \Planck\ reconstructed maps.
Here we appreciate that the noise is really the limiting factor for \Planck:
a significantly reduced noise would allow to have almost perfect 
reconstruction of bright
clusters with 
a spread reduced to 10 per cent. 
The critical issue now is how big will be the area of
 the sky where this performance
will be achieved. Because the scanning strategy and the instrument performances
are not settled yet, this is not a straightforward question to answer.
It is clear, however, that the selection function for the \Planck\ cluster
catalog 
is also going to depend on the reconstruction method and associated errors
in different areas of the sky.

\subsection{Completeness and purity of the samples}\label{sec:pco}
Once a cluster map is reconstructed, it is sensible to ask whether the 
structures found really correspond to clusters in the input map or not.
Furthermore one would like to know if a given threshold in the reconstructed
map can be associated with an input cluster intensity with high confidence.
To this aim, we tried to assess the completeness and purity of our samples 
for given output intensities.
In order to assess the purity/completeness in ACT, we smoothed the maps 
to 0.9 arcmin and considered all local maxima in the reconstructed 
map which would have a cluster counterpart in the input one within 
a radius of 0.6 arcmin.
We use the term ``purity'' to indicate the fraction of the targeted 
clusters which do have a counterpart in the input map.
We found that the purity is one  for all relevant intensities, i.e. the 
reconstruction doesn't create clusters out of nothing, at least in the 
limit of the approximations applied here (i.e. no point sources).
%\fbox{\parbox{\columnwidth}{
%
%
%
We then considered a given intensity in the input map, projected it into 
a reconstructed value using the slope calculated before and counted the 
fraction of objects that were effectively reconstructed above such threshold.
We use the term  ``completeness'' to indicate the fraction of the 
initial clusters that make that threshold in the reconstructed map. 
The sample is complete for clusters with $y$ parameter bigger than 
$3 \times 10^{-4}$ (there are about 15 such clusters in a 
$100~{\rm deg}^2$ ACT survey).
 However completeness drops to 50 per cent for $y$
above $1.5 \times 10^{-4}$ (about 150 such clusters in a $100 {\rm deg}^2$
 ACT survey).
 This is because very compact clusters are 
anyway mapped into less intense objects: they would be accounted for
if we consider the spread associated with the reconstruction. 
%
%
%
%}}

As for \Planck, we still have a very good purity level. However, the 
level of noise considered here prevents us from assessing completeness.
The dimension of the cluster (rather than its intensity) seem to be the dominant factor in selecting the clusters that are reconstructed.
This biases the reconstruction in favour of local clusters, rather than the most
intense. 
Hopefully the real noise level will be better than the one used here in 
most areas of the sky, so that a selection on the basis of brightness
rather than size will be possible.

\section{Conclusions} \label{par:con}
We investigated the issue of SZ cluster reconstruction in future
 multi-frequency CMB experiments.
We proposed a new method for component separation that is specifically 
tailored to reconstruct SZ galaxy clusters. Our approach takes into 
account that clusters produce a non-Gaussian, localized signal and
they are non-spherical, sparse objects on the sky.
The reconstruction is based in the  wavelet domain, therefore is  local and 
takes into account covariances between different scales, positions and 
orientations.
We studied the  performances of this estimator 
with different models of non-Gaussianity, some corresponding to the one
observed in SZ simulated maps and others that are common  
  in image reconstruction literature.
We show that
this method outperforms previous techniques like Wiener filtering in 
recovering the cluster central intensity both in terms of average 
reconstruction and of associated error.
This result mainly depends on the  better characterization of the bright 
pixel (corresponding to cluster's centers)
  that we achieve with  the non-Gaussian prior assumed.
The success of this strategy depends on the combined effect of using
a wavelet basis, a local reconstruction and an adequate non-Gaussian 
model for the signal.

We applied our method to two  experiments, very different in nature:
 \Planck\ and ACT.
In the case of ACT, our method
allows to recover the 80--90 per cent of the central intensity of 
our 50 brightest reconstructed
 clusters with a reconstruction error of about 10 per cent 
(mainly caused by the beam size). 
The 100 ${\rm deg}^2$ ACT survey should contain about 150
  of such clusters, and the subsample for with 
high completeness level should contain approximately 50 clusters.
This is 
an adequate sample to constrain cosmology.

In the case of    \Planck\, we were able to better
reconstruct the  most extended clusters, rather than the brightest.
  This method  allows a reconstruction of the 45 
per cent of the cluster intensity, 
improving on standard Wiener filtering by about a factor five.
The error associated with the reconstruction for the 12 targeted clusters is,
 however, still about 27
 per cent. 
 While this is an improvement on the Wiener filtering,
 it is somewhat unsatisfactory for deriving cosmology.
Another limitation derives from the particular  selection effect: most
 bright but compact clusters are not reconstructed at all.
We point out that the major limitation for \Planck\ is the noise level,
 and show that  a reduced noise level would allow almost perfect 
reconstruction with very little scatter for a sample of intensity-selected
clusters.
The area of the \Planck\ sky where this will be possible depends upon 
the final scan strategy and instrument performance, which are not precisely
defined yet. In general, we shall expect to have very different selection 
functions which will depend on sky positions: a detailed study like the 
present one will be necessary to infer actual performance.

These results may depend heavily on our neglect of point sources.  It
is important to point out, however, that the localized and non-Gaussian
nature of the point source signal will almost certainly affect the standard
Wiener-filter techniques more adversely than the non-Gaussian estimator 
presented here.  We will pursue this more in future research.

\section*{ACKNOWLEDGMENTS}

EP is an NSF-ADVANCE fellow, also supported by NASA grant NAG5-11489.
KH wishes to thank Pengjie Zhang for his assistance in providing SZ maps.

\bibliography{SZrec}
%\bibliography{SZrec,mn-jour}
\bibliographystyle{mn2e}

%\begin{thebibliography}{99}
%\bibitem[\protect\citename{Milgrom }1983c]{Milgrom83c}
% Milgrom M., 1983c, ApJ, 270, 384
%\bibitem[\protect\citename{Milgrom }1994]{Milgrom94}
% Milgrom M., 1994, Ann. Phys., 229, 384 [astro-ph/9303012]
%\bibitem[\protect\citename{Milgrom }1999]{Milgrom99}
% Milgrom M., 1999, Dark matter in astrophysics and particle physics,
% eds. H.V. Klapdor-Kleingrothaus and L. Baudis,
% Institute of Physics Pub., Philadelphia, p.$\,443$
%\end{thebibliography}

\end{document}